\documentclass[twocolumn,prd,nofootinbib,aps,floats,floatfix,amsmath,amssymb,secnumarabic]{revtex4} %
\usepackage[final]{graphicx}
\usepackage{amsmath}
\usepackage{bbm}
\usepackage{amsfonts}
\usepackage{amssymb}
\usepackage{latexsym}
\usepackage{graphicx}
\usepackage[english]{babel}
\usepackage{multirow}
\usepackage{float}
\usepackage{url}
\usepackage{hyperref}
\usepackage{slashed}
\usepackage{xcolor} 
\usepackage[utf8]{inputenc}

\def\sfrac#1#2{{\textstyle{#1\over #2}}}
\newcommand{\be}{\begin{equation}}
\newcommand{\ee}{\end{equation}}
\newcommand{\ba}{\begin{array}}
\newcommand{\ea}{\end{array}}
\newcommand{\bea}{\begin{eqnarray}}
\newcommand{\eea}{\end{eqnarray}}
\newcommand{\sss}{\scriptscriptstyle}

\newcommand{\nn}{\nonumber}
\renewcommand{\L}{{\sss L}}
\newcommand{\R}{{\sss R}}

\newcommand{\hc}{{\sss HC}}

\begin{document}

\title{\huge $B$ decay anomalies and dark matter\\ from vectorlike confinement}
\author{James M.\ Cline}
\affiliation{Niels Bohr International Academy \& Discovery Center,
 Niels Bohr Institute, University of Copenhagen,
Blegdamsvej 17, DK-2100, Copenhagen, Denmark 
and McGill University, Department of Physics, 3600 University St.,
Montr\'eal, Qc H3A2T8 Canada}

\begin{abstract}

Lepton flavor universality violating $B\to K\ell\ell$ and
$K^*\ell\ell$ decays
tentatively observed by LHCb can be explained by leptoquark exchange.
We explore a simple model for the $B$ anomalies with a composite
leptoquark from new strong dynamics at the TeV scale, a confining
SU($N_\hc$) hypercolor interaction. The new matter fields, fundamentals
under SU($N_\hc$), are  heavy vectorlike fermions $\Psi,\, S,$ and an
inert scalar doublet $\phi$.  $\Psi$ is colored under QCD while $S$ is
neutral, and the hyperbaryon $S^N$ is a dark matter candidate.   The
model is tightly constrained by meson-antimeson oscillations, lepton
flavor violation, and LHC
searches for resonant production of the exotic bound states.  The
dark matter may be detectable through its magnetic dipole moment.
If $m_S$ is sufficiently small, composite leptoquarks and
heavy lepton partners can be
pair-produced at an observable level at LHC.

\end{abstract}
\maketitle

\section{Introduction}

While the Standard Model (SM) continues to accurately describe 
particle interactions at the shortest probed distance scales, there is an
encouraging hint of new physics from decays of neutral $B$ mesons to
$K$ or $K^*$ and charged leptons.  The LHCb collaboration measures the
ratios of branching ratios \cite{Aaij:2014ora,Aaij:2017vbb}
\be
	R_K^{(*)} = {B(B\to K^{(*)} \mu^+\mu^-)\over 
	B(B\to K^{(*)} e^+e^-)}
\ee
to be $(20-30)\%$ below the SM prediction for both $K$ and $K^*$ final
states.  Uncertainties from the hadronic matrix elements
cancel in the ratios, greatly reducing the theoretical errors and
making these ratios particularly interesting probes of new physics
that could violate lepton flavor universality.   Although the
discrepancies between theory and experiment in the individual
measurements are below $3\,\sigma$, combining them increases the
significance to $4\,\sigma$
\cite{Geng:2017svp,DAmico:2017mtc,Altmannshofer:2017yso,
Capdevila:2017bsm,Celis:2017doq,Ciuchini:2017mik,Bardhan:2017xcc,Neshatpour:2017qvi,
Alok:2017jaf,Alok:2017sui,Choudhury:2017qyt}.  This conclusion
remains true when other observables are included in the fit, 
like the branching ratio for $B_s\to \phi\mu^+\mu^-$, which also has a
deficit, and angular decay distributions that are subject to hadronic
uncertainties.  

Model-independent fits to the data show that the addition of the
single effective operator
\be
	{\cal O}_{b_\L\mu_\L} = 
{c\over\Lambda^2}(\bar s_\L\gamma_\alpha b_\L)(\bar\mu_\L\gamma^\alpha \mu_\L)
\label{effop}
\ee
to the effective Hamiltonian is sufficient to give a good fit to the
observations, with \cite{DAmico:2017mtc}
\be
	{c\over\Lambda^2} = {1.0\times 10^{-3}\over{\rm TeV}^2}
\ee
The form (\ref{effop}) arises in the standard model from $W$
exchange at one loop, and can be realized in models of new
physics loop effects 
\cite{Gripaios:2015gra,Bauer:2015knc,Arnan:2016cpy,Kamenik:2017tnu,
Das:2017kfo,Kawamura:2017ecz,He:2017osj}. It 
 could also be induced by exchange of a heavy
$Z'$ vector boson, inspiring the construction of many models
of this kind \cite{Gauld:2013qba,Gauld:2013qja,Buras:2013dea,Altmannshofer:2014cfa,
Crivellin:2015lwa,Sierra:2015fma,Crivellin:2015era,Celis:2015ara,Carmona:2015ena,Fuyuto:2015gmk,Chiang:2016qov,
Kim:2016bdu,Boucenna:2016wpr,Cheung:2016exp,Boucenna:2016qad,Crivellin:2016ejn,GarciaGarcia:2016nvr,Datta:2017pfz,Ko:2017lzd,Alonso:2017bff,
Bonilla:2017lsq,Alonso:2017uky,Ellis:2017nrp,Ghosh:2017ber,Tang:2017gkz,Chiang:2017hlj,Chivukula:2017qsi,King:2017anf,Cline:2017ihf,Chen:2017usq,Baek:2017sew,
Bian:2017rpg,Dalchenko:2017shg,Megias:2017mll}.  
By Fierz rearrangement, (\ref{effop})  
takes the form suggestive of vector leptoquark exchange.  Global
analyses identify two possible vector leptoquarks $U_1$ and $U_3$
and one scalar $S_1$ as viable candidates, in the notation of 
ref.\ \cite{Dorsner:2016wpm} where the subscript denotes the SU(2)$_L$
representation.  Models involving leptoquark exchange have been 
studied in refs.\ \cite{Hiller:2014yaa,Gripaios:2014tna,Allanach:2015ria,Alonso:2015sja,Bauer:2015knc,
Sahoo:2015pzk,Kumar:2016omp,Das:2016vkr,Li:2016vvp,Chen:2016dip,Becirevic:2016yqi,Becirevic:2016oho,
Mileo:2016zeo,Hiller:2016kry,Sahoo:2016pet,Popov:2016fzr,Barbieri:2016las,
Cline:2017lvv,Chen:2017hir,Crivellin:2017zlb,Hiller:2017bzc,
Cai:2017wry,Becirevic:2017jtw,Matsuzaki:2017bpp,Chauhan:2017ndd,
Altmannshofer:2017poe,Buttazzo:2017ixm,Crivellin:2017dsk,Guo:2017gxp,Aloni:2017ixa,Assad:2017iib,DiLuzio:2017vat,Calibbi:2017qbu}.

In the present work we offer a simple, UV complete model in which the
vector $U_1$ leptoquark arises from confining dynamics of an
SU($N_\hc$) gauge theory, referred to as hypercolor.  It is an example
of vectorlike confinement \cite{Kilic:2009mi} in which the 
constituent fermions have bare masses that can be freely chosen.  Our proposal
differs from  previous models for $R_{K^{(*)}}$ with composite
leptoquarks \cite{Barbieri:2016las,Gripaios:2014tna,Matsuzaki:2017bpp,Buttazzo:2017ixm} where the constituent masses were assumed to be
much lighter than the confinement scale $\Lambda_\hc$.  In that case,
approximate chiral symmetry is important, and the lightest bound
states are pseudo-Nambu-Goldstone bosons (pNGBs).  

Here we instead assume that
the constituent masses $M$ are larger than $\Lambda_\hc$ (but not too
much larger) so that chiral dynamics plays no role, yet the system
is not quirky \cite{Kang:2008ea} as occurs if $\Lambda_\hc \ll M$.  One
advantage of this choice is that the strong dynamics is somewhat more
calculable, via a nonrelativistic potential model that works
reasonably well for QCD -onium states like $J/\psi$ and $\Upsilon$.
Moreover collider constraints on production of the new bound states
are under better control in this regime, where resonant production
is suppressed by the wave function of the constituents, and pair
production is limited by the large bound state masses.  We also thereby avoid
complications associated with composite Higgs scenarios.

Another distinctive feature of our model is that it provides a
dark matter candidate, which is intimately linked to the $B$ decay
phenomenology.  In particular, one of the hyperquarks (denoted by $S$)
is a singlet under the SM gauge group.  It is a constituent of the
composite leptoquark, and the baryon-like $S^{N_\hc}$ bound state is
the dark matter.  Exceptionally, $S$ can be lighter than $\Lambda_\hc$
without entailing a pNGB, since the
approximate U(1)$_A$ flavor symmetry associated with $S$ is anomalous.

A further difference relative to previous models of composite
leptoquarks is that we invoke a scalar hyperquark $\phi$, a  doublet
under SU(2)$_L$.  It is motivated by the need to couple  the new
physics to left-handed quarks and leptons, eq.\ (\ref{effop}).
Previous models achieved this by taking the fermionic hyperquarks to
be doublets, but it is more straightforward to have a dark matter
candidate if this is avoided, and the massless limit of $S$ can be
safely taken without introducing new relatively light scalars that would be
easily produced in colliders.  In our model both fermionic hyperquarks
$S,\Psi$, are isosinglets, while $\Psi$ carries color. The $S\bar\Psi$
bound state is the leptoquark.

After defining the model in section \ref{model_sect} and 
reducing it to an effective theory suitable for addressing 
the $B$ decay anomalies in section \ref{EFT}, 
we identify (without fully delineating) some
regions of parameter space that are consistent with constraints
on flavor changing neutral currents (FCNCs) in 
section \ref{flavor_sect},
dark matter direct detection (section \ref{dm_constraints}),
and collider searches (section \ref{collider}).  The examples we focus
on have potential for associated new signals in all of
these categories of observables.  A summary and conclusions are
given.  Appendices provide details of the potential model for estimating
bound state properties, the formalism for computing dipole moments
from partial compositeness of the light fermions, and the magnetic
dipole moment of the dark matter constituent fermion.

\begin{figure}[t]
\hspace{-0.4cm}
\centerline{
\includegraphics[width=0.95\hsize]{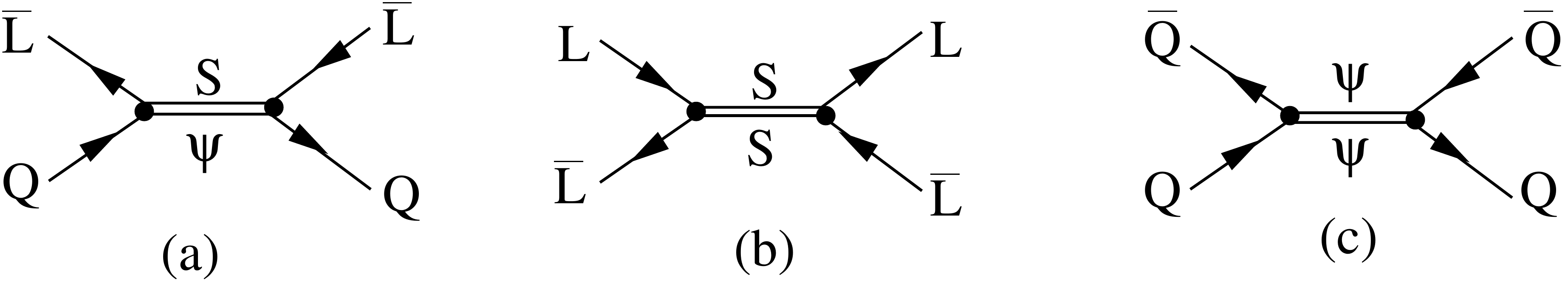}}
\caption{Tree-level contributions to flavor-violating processes
from exchange of composite vector bosons.}
\label{tree}
\end{figure}

\section{Model}
\label{model_sect}

Our model introduces three matter fields that are fundamentals
of hypercolor (HC): a Dirac fermionic DM particle $S$, 
a vectorlike fermion $\Psi$ that carries SM color, and a
scalar $\phi$ that is an SU(2)$_L$ doublet.  The quantum numbers are
shown in table \ref{tab1}.
Gauge symmetry allows these fields to couple to left-handed SM 
quarks and leptons only through the interactions 
\be
{\cal L} = \tilde\lambda_i\, \bar Q_{i,a}\, \phi^a_{ A}
\Psi^{ A} + 
	\lambda_i\, \bar S_{ A}{\phi_{ a}^{* A}}\, L_i^a
\label{Lint}
\ee
where $a (A)$ is the SU(2)$_L$ (SU($N$)$_\hc$) index and $i$ is the
flavor index.  These interactions explicitly break the approximate
flavor symmetries of the standard model (SM), and so one can
anticipate that the matching of the observed $b\to s\mu^+\mu^-$ FCNC
will give rise to other FCNC observables. The observed
$B$ decay anomaly is mediated by tree-level exchange of a
composite leptoquark $\Phi = \Psi\bar S$,  while the other kinds of
bound states  $\Psi\bar\Psi$ and $S\bar S$ give analogous 
contributions to neutral meson mixing and flavor-violating decays of
charged leptons, depicted in fig.\ \ref{tree}.

To fully specify the model, we must define the flavor basis
of the SM fields in (\ref{Lint}).  For simplicity we  assume that
the mass matrices of the charged leptons and down-type quarks are
already diagonal in this basis, so that all the mixing giving rise to
the CKM matrix is due to diagonalizing the up-type quark masses.
Then going to the mass eigenbasis of the quarks, 
\bea
	\tilde\lambda_i \bar Q_i &\to& 
	\tilde \lambda_j\left(\begin{array}{cc} \bar u_{\L,i}
	V_{ij}, &
	\bar d_{\L,j} \end{array}\right)\nn\\
	&\equiv& \left(\begin{array}{cc} \tilde\lambda'_i \bar u_i,
	& \tilde\lambda_i \bar d_i \end{array}\right)
\label{CKMmix}
\eea
Thus the couplings to up-type quarks are given by 
$\tilde\lambda'_i = V_{ij}\tilde\lambda_j$.

It is worth noting that the individual lepton flavor symmetry
for generation $i$ becomes exact in the limit $\lambda_i\to 0$:
these couplings are only multiplicatively renormalized, up to 
neutrino mass insertions.  Therefore if we only wish to explain 
lepton flavor universality violation in the muon sector, it is
technically natural to set the couplings $\lambda_1, \lambda_3 \to 0$.

If $S$ is lighter than $\phi$ and $\Psi$,  the hyperbaryon bound state
$\Sigma = S^N$ is a dark matter candidate.  Its stability is protected
by the accidental $Z_2$ symmetry under which all the new matter fields
are odd.  It is a consequence of the SU($N_\hc)$ gauge
symmetry, analogous to baryon number conservation in the SM. 
Moreover the fields
$\Psi$ and $S$ can consistently be assigned normal baryon and lepton
number, respectively.

\begin{table}[t]
\begin{tabular}{|c|c|c|c|c|c|c|}
\hline
 & SU(3) & SU(2)$_L$ & U(1)$_y$ & U(1)$_{\rm em}$ & SU(N)$_\hc$ &
$Z_2$\\
\hline
$\Psi$ & $3$ & $1$ & $2/3$ & $2/3$ & $N$ & $-1$\\
$S$    & $1$ & $1$ & $0$ & $0$ &  $N$ & $-1$\\
$\phi$ & $1$ & $2$ & $-1/2$ & $(0,-1)$ &  $\bar N$ & $-1$\\
\hline
\end{tabular}
\caption{Quantum numbers of new physics particles}
\label{tab1}
\end{table}

\section{Low-energy effective theory}
\label{EFT}

The $\bar\Psi S$ bound states of our model have the quantum numbers of
leptoquarks and contribute to the decays $B\to K^{(*)}\mu^+\mu^-$.
The leptoquarks can be either pseudoscalar $\Pi$ or vector $\rho_\mu$,
with decay constants
\bea
	\langle 0|(\bar S \gamma^\mu\gamma_5\Psi)|\Pi\rangle &=& 
	f_\Pi\, p_\Pi^\mu \nn\\
\label{vec_me}
	\langle 0|(\bar S \gamma^\mu\Psi)|\rho_\lambda\rangle &=& 
	f_{\rho}\,m_{\rho}\epsilon_\lambda^\mu
\label{decay_const}
\eea
where $\lambda$ labels the helicity state of the vector. 
These currents couple to the SM fields via $\bar Q\gamma_\mu L$ and
so the interactions of the pseudoscalar are suppressed by small quark and
lepton masses through the equations of motion, due to the momentum
factor $p_\Pi^\mu = p_q^\mu + p_l^\mu$.  This can also be understood in terms of the 
helicity suppression of the amplitude for pseudoscalar decay to
approximately chiral states in 
analogy to charged pion decay.  
Hence we are interested in the vector leptoquark
for explaining the $B$ decay anomalies.

We seek an effective description of the leptoquark interaction with
the SM fields,
\be
	g_{\rho}^{ij} \rho_\mu\, (\bar Q_i\gamma^\mu L_j)
\label{Lint2}
\ee
The coupling $g_{\rho}^{ij}$ can be determined by matching the 
decay rate for $\rho_\mu\to Q_i \bar L_j$ in the effective theory and
the underlying model.  In the effective theory, the rate is 
\be
	\Gamma(\rho_\mu\to Q_i \bar L_j) = {|g_{\rho}^{ij}|^2\over
24\pi} m_{\rho}
\ee
neglecting the quark and lepton masses.  
In the UV theory, this rate can be computed as $\Gamma = \sigma
v_{\rm rel}|\psi(0)|^2$, where $\psi$ is the wave function for the 
$\Psi\bar S$ bound state and $\sigma$ is the cross section for
$\Psi\bar S\to Q_i\bar L_j$ annihilation,
\be
	\sigma v_{\rm rel} = N_\hc{|\tilde\lambda_i^2\lambda_j^2| (m_S +
	m_\Psi)^2\over 96\pi (m_S\, m_\Psi + m_\phi^2)^2}
\ee
ignoring the initial velocities of the bound particles, and assuming
only the spin-1 configuration of $S$ and $\Psi$ contributes in the
sum over spins.  This gives
\be
	g_{\rho}^{ij}  = \left(N_\hc\over 4\,m_\rho\right)^{1/2}
	{\tilde\lambda_i \lambda_j^* (m_S +
	m_\Psi)\over  (m_S\, m_\Psi +
m_\phi^2)}\,\psi(0)
\label{grho_eq}
\ee

Once the effective interaction (\ref{Lint2}) is specified, we can
integrate out the leptoquark to generate the 
dimension-6 operator shown in fig.\ \ref{tree}(a),
\bea
	\delta{\cal L}&=& -{g_{ij}g^*_{i'j'}\over
m_\rho^2}(\bar Q_i^a\gamma^\mu Q_{i',b})
	(\bar L_{j'}^b \gamma^\mu L_{j,a})
\label{dL1}
\eea
after Fierz transforming \cite{Nishi:2004st}, where $a,b$ are
SU(2)$_L$ indices.  This contribution must interfere destructively
with the SM contribution to $b\to s\mu^+\mu^-$,
requiring $g_{32}^*g_{22}$ to be
approximately real and positive.

In addition to the composite leptoquarks, there are composite
$\bar\Psi\Psi \equiv \rho_\Psi$ and $\bar S S \equiv\rho_S$ 
vector bosons, whose exchanges are shown in
 fig.\ \ref{tree}(b,c).   The
ensuing operators are
\bea
\delta{\cal L}= && -{h_{ij}h^*_{i'j'}\over 2\,m_{\rho_S}^2}
	(\bar L_i^a\gamma^\mu L_{i',b})
	(\bar L_{j'}^b \gamma^\mu L_{j,a})\nn\\
	&-& {k_{ij}k^*_{i'j'}\over 2\,m_{\rho_\Psi}^2}
	(\bar Q_i^a\gamma^\mu Q_{i',b})
	(\bar Q_{j'}^b \gamma^\mu Q_{j,a})
\label{dim6op}
\eea
where $h_{ij}$ and $k_{ij}$ 
are determined analogously to 
(\ref{grho_eq}), with $\tilde\lambda_i \lambda_j^*\to
\lambda_i \lambda_j^*$ and $\tilde\lambda_i \tilde\lambda_j^*$,
respectively.

\section{Flavor constraints}
\label{flavor_sect}

The fit of ref.\ \cite{DAmico:2017mtc} implies that the $R_{K^{(*)}}$ anomalies can
be explained by taking
\be
	 {g_{22}\, g_{32}^*\over m_\rho^2} = {1\times 10^{-3}\over {\rm
TeV}^2}
\label{fiteq1}
\ee
To translate this into a constraint on parameters of the model, we
must determine $\psi(0)$ and the bound state mass.  For this purpose
we use a potential model that is described in appendix \ref{app3}.
To simplify this
preliminary analysis, we assume that $m_\Psi\cong
m_\Phi$.\footnote{Also this choice maximizes the wave function at the 
origin}\ \  
We consider two possibilities for the dark constituent mass,
$\Lambda_\hc \lesssim m_S\lesssim m_\phi \equiv M$ and $m_S \ll \Lambda_\hc
\lesssim m_\phi$, such that $S$ is respectively nonrelativistic or
relativistic within the bound state.  Two version of the potential
model are given to treat these cases.  We avoid the quirky regime where
$\Lambda_\hc\ll M$ \cite{Kang:2008ea} since collider constraints are
expected to become more stringent (and difficult to quantify), as we
will discuss in section \ref{collider}. 

\begin{figure}[t]
\hspace{-0.4cm}
\centerline{
\includegraphics[width=0.95\hsize]{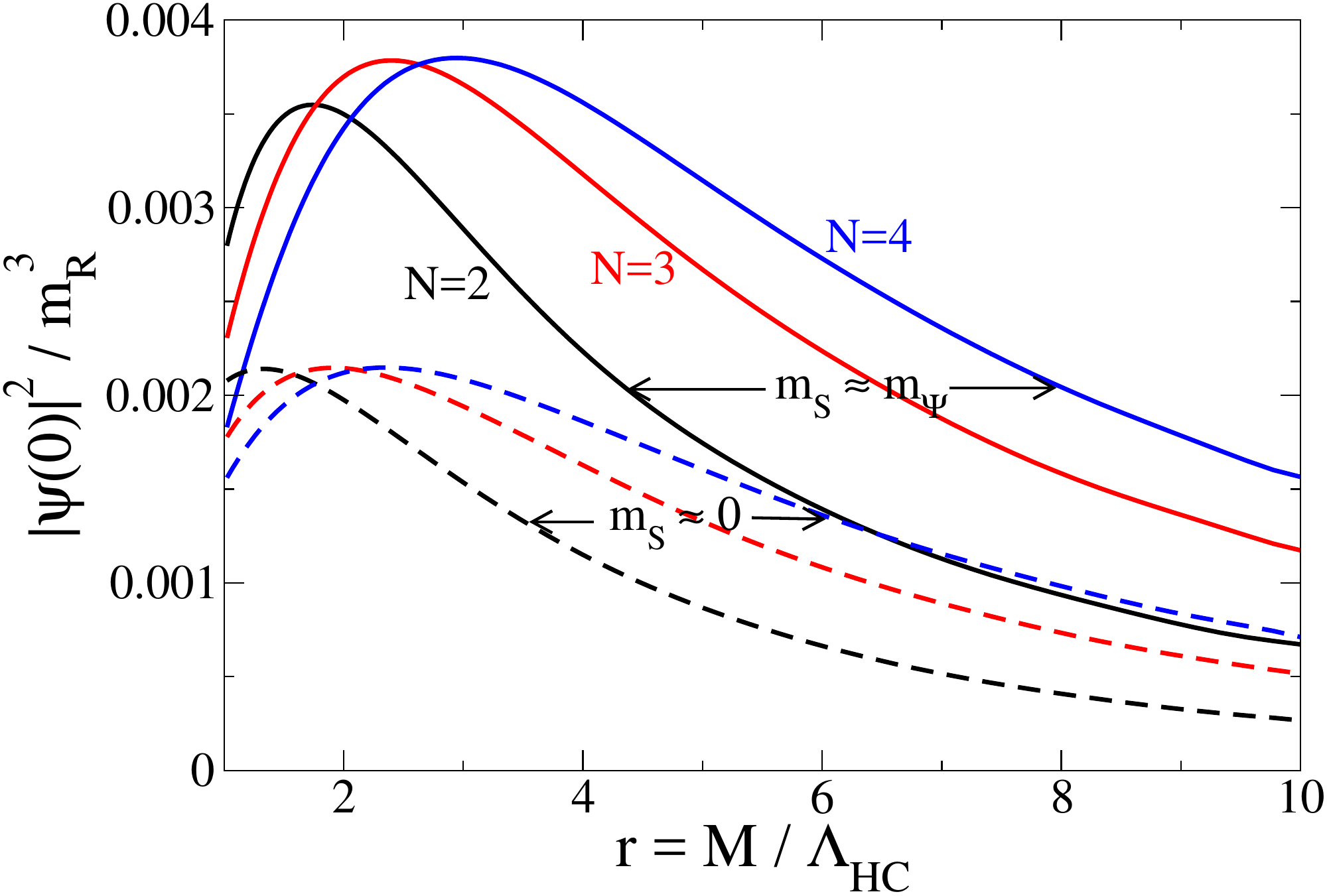}}
\caption{The function $\zeta = |\psi(0)|^2/m_R^3$ (eq.\ (\ref{zeq}))
that encodes the strong dynamics dependence of the exchange of bound
state resonances between SM fermion bilinears.  Solid curves are
for all constituents having same mass $M$, while dashed are for the
case of $m_S\ll \Lambda_\hc$.}
\label{zetaf}
\end{figure}

In either case, the coefficients of the dimension-6 operators
(\ref{dim6op}) are proportional to
\be
 	\zeta \equiv{|\psi(0)|^2\over m_R^3}
\label{zeq}
\ee
that encodes the effects of the confining dynamics, where $m_R$ is the mass of
the exchanged resonance.  $\zeta$ is dimensionless and thus only 
depends upon ratios of mass or energy scales (and $N_\hc$).  In the case where all
masses are approximately equal to scale $M$, there is just one such 
ratio, $r = M/\Lambda_\hc$.  Using the nonrelativisitic potential model in
the region $M/\Lambda_\hc > 1$, we numerically determine $\zeta(r)$
for $N_\hc = 2,3,4$, shown in fig.\ \ref{zetaf} (solid curves).  It reaches a maximum
value of $\sim 0.0037$ for $r\cong 3$ if $N_\hc = 3,4$, and is smaller
if $N_\hc = 2$.  For the case $m_S\ll \Lambda_\hc$, the low scale 
$m_S$ becomes
irrelevant and we can can still express $\zeta$ as a function of the
same $r = M/\Lambda_\hc$, using the relativistic version of the
potential model.  $\zeta$ is smaller for these models, $\sim 0.002$
for $r\sim (2-3)$.

Since $\zeta$ is small, the best case for avoiding
large couplings $\lambda_2$, $\tilde\lambda_{2,3}$ in the underlying
model is to have $m_\phi\sim m_\Psi \sim M$
and $\Lambda_\hc\sim (0.3-0.4)M$ for $N_\hc = 3$ or 4.  Then with
$\zeta = 0.0037$, eq.\ (\ref{fiteq1}) implies
\be
	|\lambda_2^2\,\tilde\lambda_2\,\tilde\lambda_3|
	= 0.3\left(M\over{\rm TeV}\right)^2 \left(3\over N_\hc\right)
\label{fiteq2}
\ee
From the potential model, the bound state mass $m_\rho$ at $\Lambda/M = 0.4$ and $N_\hc = 3$ is 
predicted to be 1.75 times the naive value of $2 M$  that ignores
the contribution from the hypercolor flux tube.  For $M=1\,$\,TeV
therefore, $m_\rho = 3.5\,$TeV.

In the case where $m_S\ll\Lambda_\hc$, the couplings must be increased
by $\sim$$15\%$, to compensate for the smaller $\zeta$
factor.  From the partially  relativistic version of the potential
model, the mass of the resonance is found to be approximately
\be
	m_\rho \cong \left(0.56+ 1.8\, N_\hc\right) \Lambda_\hc + m_\Psi
\ee
valid for $m_\Psi/\Lambda_\hc\gtrsim 3$; for smaller values $m_\rho$
tends toward a constant independent of $m_\Psi$.
	
\subsection{Meson-antimeson mixing}
\label{mam}

Diagrams of the type fig.\ \ref{tree}(c) contribute to the
oscillations of neutral mesons: $K^0$-$\bar K^0$, $D^0$-$\bar D^0$,
$B^0$-$\bar B^0$ and $B_s^0$-$\bar B_s^0$, placing constraints
on different combinations of the $\tilde\lambda_i$ couplings by
similar reasoning as led to eq.\ (\ref{fiteq2}).  For the fiducial
model parameters with $\zeta = 0.0037$ and $N_\hc = 3$, we
find the upper limits given in table \ref{tab2}.  These are inferred
by comparing the operator coefficients $k_{ij}^2/2 m_{\rho_\Psi^2}$ with
the upper limits from ref.\ \cite{Arnan:2016cpy} for $B_s$ mixing
and refs.\ \cite{Bona:2007vi,Bona:2016bvr} for $K$, $D$ and $B_d$
mixing (which gives the most stringent bounds in each case).
The bound on $K$ and $D$ mixing are conservative in the sense that we have
assumed that the contribution to the imaginary part of the amplitude,
which is much more strongly constrained, is
as large as that of the real part. 

As an example of parameters that can satisfy these constraints, 
we take $M=1$\,TeV and 
\bea
	\tilde\lambda_1 &=& -0.01,\quad
	\tilde\lambda_2 = 0.1,\quad
	\tilde\lambda_3 = 0.66,\quad 
	\lambda_2 = 2.1\nn \\
	(\tilde\lambda_1' &=& 0.014,\quad
	\tilde\lambda_2' = 0.13,\quad
	\tilde\lambda_3' = 0.66)
\label{params}
\eea
The primed couplings (pertaining to the up-type quarks) are determined
by the unprimed ones through eq.\ \ref{CKMmix}.  The resulting
contributions to the products of couplings relevant to meson mixing
are shown in table \ref{tab2} (last column).  
The relative minus sign between $\tilde\lambda_1$ and
$\tilde\lambda_{2}$
 leads to an accidental cancellation in 
$V_{ij}\tilde\lambda_j$ such that the $D^0$ mixing contribution is
just below the limit; without this sign, we would get
$\tilde\lambda_1'\tilde\lambda_2' = 0.004$, saturating the limit.

The fact that all the mixing constraints are close to being saturated,
with the exception of $B_s$, can be understood from the lack of any
approximate flavor symmetry such as minimal flavor violation 
\cite{DAmbrosio:2002vsn} in this model.  By requiring a large enough $b\to s\mu^+\mu^-$
transition and trying to avoid couplings much larger than 1, we are pushed
into this restricted region of parameter space.

One way to relieve the
tension is to take smaller values of $M$, say 800\,GeV.  Then the
couplings could on average be smaller by a factor of $\sqrt{0.8}$
than in eq.\ (\ref{params}).   If we keep the leptonic coupling 
large, $\lambda_2 = 1.1$, the quark couplings can all be reduced by 
$0.8$, and the last column in table \ref{tab2} is reduced by
$0.64$, leading to a $60\%$ reduction in the actual amplitude.
Allowing for larger leptonic coupling can further ameliorate the
situation, which is thus not as marginal 
as the limiting case makes it appear.  
$M$ cannot be made lower than 800\,GeV because of
LHC constraints, as we will see in section \ref{collider}.
Our main goal in this paper is to establish some region of viability
for the model.  A thorough exploration of the parameter space 
will become more strongly motivated if the $B$ decay anomalies are
confirmed with greater significance.

\begin{table}[t]
\begin{tabular}{|c|c|c|c|}
\hline
meson & quantity &$\hbox{ upper limit}\atop\hbox{ (units $M$/TeV)}$ 
& $\hbox{fiducial value}\atop\hbox{ (units $M$/TeV)}$\\
\hline
$K^0$ & $|\tilde\lambda_1\tilde\lambda_2|$ & $1.3\times
10^{-3}$ & $1\times 10^{-3}$ \\
$D^0$ & $|\tilde\lambda_1'\tilde\lambda_2'|$ & $2\times 10^{-3}$ & 
$7\times 10^{-4}$\\
$B^0$ & $|\tilde\lambda_1\tilde\lambda_3|$ & $0.026$ & $0.0066$\\
$B_s^0$ & $|\tilde\lambda_2\tilde\lambda_3|$ & $0.066$ & $0.066$\\
\hline
\end{tabular}
\caption{3rd column: bounds from meson-antimeson mixing on quark couplings.
$M$ is the (assumed) universal mass of the constituents.
Last column: values following from fiducial parameter choices,
eq.\ (\ref{params}).}
\label{tab2}
\end{table}

\subsection{Lepton flavor violating decays}

As mentioned above, nothing obliges us to turn on the couplings
$\lambda_1$, $\lambda_3$ to $e$ and $\tau$ leptons in this model,
apart from aesthetic considerations.  If these couplings are
nonvanishing, then the diagram of fig.\ \ref{tree}(b) contributes to 
lepton-flavor violating (LFV) decays $\tau\to 3\ell$ (where $\ell$ is
$\mu$ or $e$) and $\mu\to 3e$.  We can derive upper bounds on 
$|\lambda_1|$, $|\lambda_3|$ by comparing the amplitude from
exchange of the $\rho_S = S\bar S$ vector meson,
\be
	\sfrac34\zeta\,{\lambda_i\lambda_j^*|\lambda_j|^2\over 2M^2}
	(\bar\ell_j\gamma^\mu\ell_i)(\bar\ell_j\gamma^\mu\ell_j)
\ee
to the SM amplitude $2\sqrt{2}\,G_F 
(\bar\nu_i\gamma^\mu\ell_i)(\bar\ell_j\gamma^\mu\nu_j)$ for the 
corresponding allowed decay, giving a
ratio of branching ratios of $R_{ij} = 1.8\times 10^{-9}|\lambda_i|^2
|\lambda_j|^6$.  From $\tau\to 3\mu$ we require $R_{\tau\mu} < 1.2\times
10^{-7}$ and from $\mu\to 3e$, $R_{\mu e} < 10^{-12}$, 
giving 
\be
	|\lambda_1| < 0.23,\quad |\lambda_3| < 0.9
\label{lfvdecay}
\ee	
This implies that 
the anomalous contribution to the $b\to s e^+e^-$ 
amplitude (which goes
as $|\lambda_1|^2$) is less than 0.01 times that of 
$b\to s \mu^+\mu^-$.

%
\begin{figure}[t]
\hspace{-0.4cm}
\centerline{
\includegraphics[width=0.95\hsize]{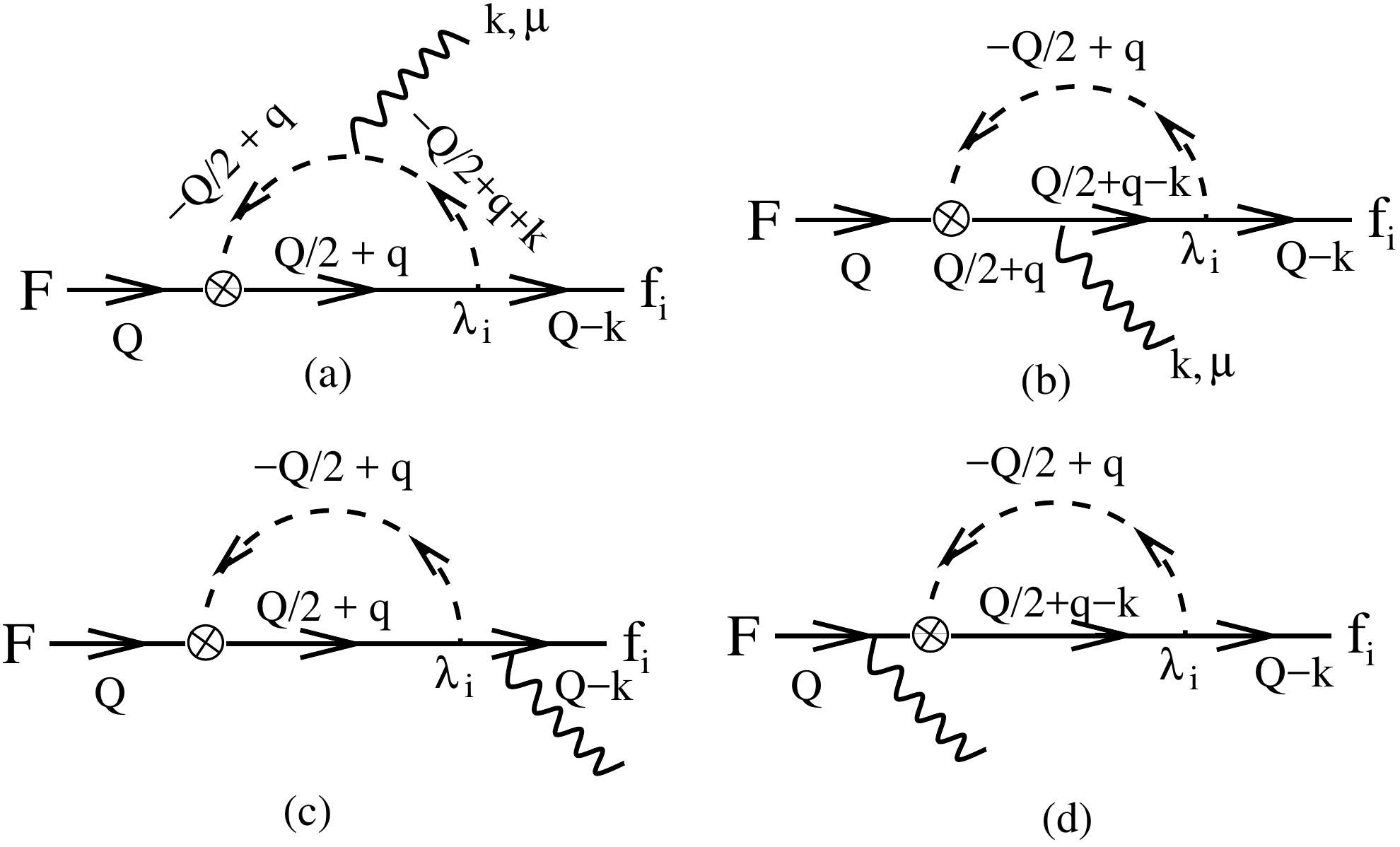}}
\caption{Diagrams for transition magnetic moment from heavy
composite to light fundamental fermion. Equal constituent masses
are assumed for routing of momenta. Cross represents the
bound-state wave function.}
\label{bound-state}
\end{figure}
%

\bigskip
\subsection{Radiative transitions}

Another potentially important class of FCNCs is the radiative
decays such as $b\to s\gamma$, $\tau\to \mu\gamma$.  They
arise through the mass mixing of SM fermions with heavy composite
fermions from the confining sector.  
These are vectorlike doublet leptons and quarks,
$\phi S \equiv F_\ell$ and $\phi\Psi\equiv F_q$,
that naturally have magnetic moments coming from their charged
constituents.  They contribute to dipole operators of the light fermions
in the mass eigenbasis, including transition magnetic moments.

The mass mixing in this kind of model was previously considered 
in ref.\ \cite{Cline:2016nab}.  Here we make the connection to
transition moments more precise, using methods that were previously
applied to the rare decays of $Z\to\Upsilon\gamma$ in the SM
in ref.\ \cite{Guberina:1980dc}.  The contributions look like
loop diagrams (fig.\ \ref{bound-state}) but the integral is over the
relative momentum of the bound constituents and is weighted by
the wave function in momentum space.  Details are given in appendix
\ref{app2}.   There we show that the mass-mixing between the SM 
and heavy fermions takes the form
\bea
\label{comp_mass}
	\tilde\lambda_f \bar Q_{f,a} \phi^a \Psi + 
	\lambda_i \bar S\phi_a^* L_i^a 
	&\to &
	{\psi(0)\over\sqrt{M}}\left(\tilde\lambda_i \bar Q_i F_q  +
	\lambda_i \bar F_\ell L_i\right)\nn\\
	&\equiv& \mu_{q}^{i}\bar Q_i F_q + \mu_{\ell}^{i} 
	\bar F_\ell L_i
\eea
where $\psi$ is the wave function of the bound state and $M$
is the mass of the constituents, that we continue to take to be equal
for simplicity.  For unequal masses, $M$ is replaced by twice the
reduced mass.

In addition, as shown in appendix \ref{app2}, the diagrams of
fig.\ \ref{bound-state} give rise to 
dimension-6 transition moment operators of the form
\be
{\cal L} = {-e F^{\mu\nu}\over 2M_F^2}\left(
q_q\,\mu_{q}^{f}\,\bar Q_f \sigma_{\mu\nu} F_q + q_\ell\,\mu_{\ell}^{f} \,\bar F_\ell
\sigma_{\mu\nu} L_f\right)
\ee
where $q_f$ is the electric charge of particle $f$, a quark or lepton.  
Ref.\ 
\cite{Esposito:1985su} noted that these lead to transition magnetic moments
among the light fermion states upon diagonalization of the mass
matrices, which take the form
\be
\left(\begin{array}{cc} \bar f_\R & \bar F_\R\end{array}\right)
\left(\begin{array}{cc} m_f & 0\\ \mu_f & M_{F}\end{array}\right)
\left(\begin{array}{c} f_\L \\ F_\L\end{array}\right)
\label{mmat}
\ee
where $f,F$ denote the SM and heavy fermions, respectively, with
$m_f$ being the SM mass matrix  and $M_F$ the heavy composite mass.
To leading order for small mixing, (\ref{mmat}) is diagonalized by separate left- and right-handed 
transformations ${\cal O}_{\L,\R} = \left({1\atop-\theta_{\L,\R}^T}
{\theta_{\L,\R}\atop 1}\right)$
with mixing angles that are vectors in flavor space,
\be
 \theta_\R = {m_f \mu_f\over M_{F}^2},\quad
\theta_\L = {\mu_f\over M_{F}}
\ee
After transforming to the mass eigenbasis, a transition moment for
the light fermions is generated,
\be
{\cal L} = 	e {q_f \mu_f^i\mu_f^j m_f^j\over  2M_F^4}
	(\bar f_{\L,i}\,\sigma_{\mu\nu} f_{\R,j}) F^{\mu\nu}
\label{trans_moment}
\ee
It is proportional to $|\psi(0)|^2/M_F^4 = \zeta/M_F$ 
rather than just $\zeta$ (recall eq.\ (\ref{zeq})).  The same potential
model can be used to compute $M_F$ as we used for the bosonic
bound states, since quantum mechanics
is insensitive to the spins of the constituent particles, and 
we are ignoring the spin-spin interactions in our simple treatment.
Then for the fiducial parameters with $\Lambda_\hc = 0.4\,M$,
$|\psi(0)|^2/M_F^4 = \zeta/(3.5\,M)$.

%
\begin{figure*}[t]
\hspace{-0.4cm}
\centerline{
\includegraphics[width=0.5\hsize]{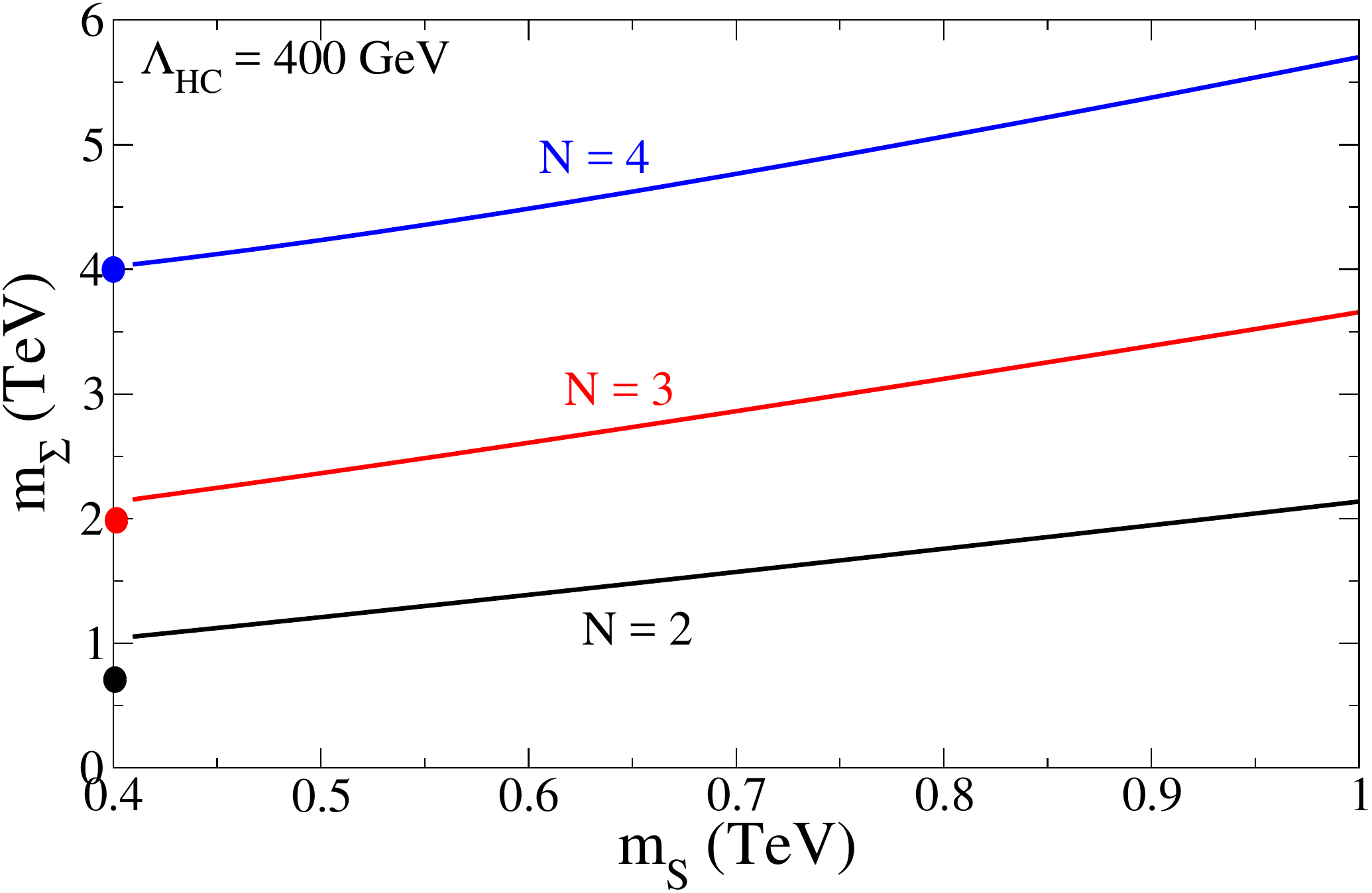}\includegraphics[width=0.47\hsize]{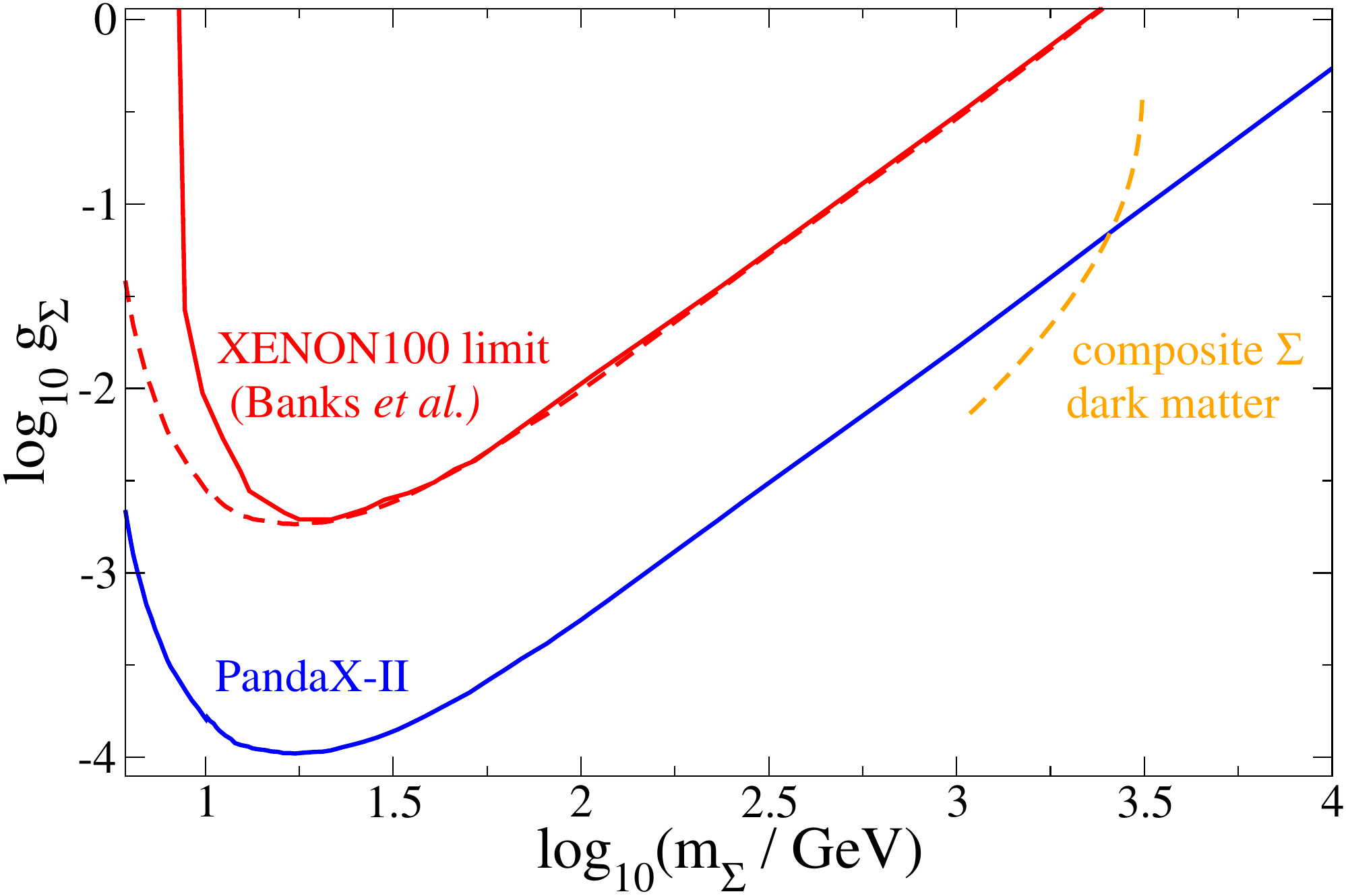}}
\caption{Left: hyperbaryonic dark matter mass $m_\Sigma$ versus
constituent mass $m_S$ for $N_\hc = 2,3,4$ and $\Lambda_\hc = 400$\,
GeV.  Heavy dots show the value of $m_\Sigma$ in the limit
$m_S\to 0$.  Right: predicted gyromagnetic moment for the $N_\hc=3$ 
model (orange dashed) and current
constraint from direct detection \cite{2017arXiv170806917P} (solid
blue), based on rescaling of earlier limit from 
ref.\ \cite{Banks:2010eh} (solid red).  Dashed red is an estimate of 
the improved relative sensitivity at low mass in recent experiments, compared to
the XENON 100 \cite{Aprile:2010um} limit.}
\label{dirdet}
\end{figure*}
%

\subsubsection{FCNC quark transitions}

From the general formula (\ref{trans_moment}), the operator 
contributing to $b\to s\gamma$ is
\be
	{e\tilde\lambda_2\tilde\lambda_3|\psi(0)|^2 m_b\over 6
M_{F_d}^4\, M}\,
	\bar s_\L \slashed {q}\gamma^\mu b_\R
\label{b2sg_amp2}
\ee
This implies that the conventionally normalized 
Wilson coefficient of the dipole operator ${\cal O}_7$ is
\be
C_7 = {\sqrt{2}\pi^2\tilde\lambda_2\tilde\lambda_3\,\zeta
\over 10.5\, G_F V_{ts} V_{tb} M^2 }\cong 7\times 10^{-4}
\ee
using the parameter
values (\ref{params}) with $\Lambda_\hc =400$\,GeV, well below
the
experimental limit of $\sim 0.02$ \cite{Descotes-Genon:2015uva}.  
Under the scaling of parameters
described at the end of section \ref{mam}, this prediction
remains unchanged.

Similarly for $c\to u\gamma$, we predict a transition magnetic moment
like (\ref{b2sg_amp2}) with $\tilde\lambda_i\to\tilde\lambda_i'$,
eq.\ (\ref{CKMmix}).  Using the values (\ref{params}), the
contribution to the $c\to u\gamma$ amplitude is $\sim 10^{-2}$ 
smaller than
the SM contribution.  Likewise, the contribution to $d\to
s\gamma$ is orders of magnitude below the limit from
$K_S\to\pi^0\ell^+\ell^-$ decays (see section 7.7 of \cite{Cline:2015lqp}).

\subsubsection{FCNC leptonic decays}
The radiative decays $\ell_i\to\ell_j\gamma$ proceed
analogously to those of the quarks, due to mixing with the heavy
composite lepton whose mass is 
$M_{F_\ell} \cong 3.5\, M$ for our fiducial parameter choice.  
The induced transition
magnetic moment is 
\be
\mu_{ij} = {e\lambda_i\lambda_j \zeta\,m_i \over M_{F_\ell} M}
\ee
and the partial decay width is 
$\delta\Gamma = \mu_{ij}^2
m_i^3/8\pi$ \cite{Giunti:2014ixa}.  This gives limits
\be
	|\lambda_1| < 7.5\times 10^{-4},\quad |\lambda_3| < 0.56
\ee
respectively, from $\mu\to e\gamma$  and $\tau\to\mu\gamma$
that are more stringent than (\ref{lfvdecay}).

\subsubsection{Muon anomalous magnetic moment}
The contribution to the muon anomalous magnetic moment is
\be
	a_\mu = {(g-2)\over 2} = |\lambda_2|^2{ m_\mu^2 \,\zeta
	\over M_{F_\ell}\,M}
\ee
For our benchmark values,
we get $a_\mu \sim 10^{-11}$, which is 300 times too small 
\cite{PDG1}.  It will be seen below that LHC constraints prevent
taking $M_{F_\ell}$ to be smaller than the TeV scale, leaving 
little room for increasing $a_\mu$.

\section{Composite dark matter}
\label{dm_constraints}

 The $\eta'$-like $S\bar S$ bound state is unstable to
self-annihilation into $\mu^+\mu^-$ through $\phi$ exchange,
but the hyperbaryonic bound state $\Sigma \equiv S^N$ is a  stable dark
matter candidate.  Determination of its thermal relic density
can depend upon the annihilation of the constituent $S\bar S$
particles
above the SU($N_\hc$) confinement transition, as well as 
$\Sigma\bar\Sigma$ annihilations following confinement.  
For the para-meters of interest, with
$\Lambda_\hc\sim 400$\,GeV and $m_S \lesssim 1$\,TeV, 
it has been shown 
\cite{Cline:2016nab,Mitridate:2017oky} that the resulting  relic
density is too small (by a factor of $\sim 1000$).
However since hyperbaryon number is conserved,\footnote{note that
$S$, $\Psi$ and $\phi^*$ can be assigned hyperbaryon number $+1$} it is possible that an
asymmetry between $S$ and $\bar S$ was produced in the early universe,
so that $\Sigma$ can be asymmetric dark matter.  We make this
assumption here, without attempting to account for the origin of the
asymmetry. 

From the nonrelativistic potential model, the DM mass can be
numerically determined as a function of $m_S$.   
In the range $1 < 
m_S/\Lambda_\hc < 10$, we find the approximate fit
\be
	m_\Sigma \cong \left(3.8(1-N_\hc) + N_\hc^2\right)\Lambda_\hc
 + \left(0.3+0.8\,N_\hc\right)m_S
\label{msigma}
\ee
The dependence on $m_S$ is shown in fig.\ \ref{dirdet}(left),
including the limit $m_S\to 0$ (heavy dots on $y$-axis), obtained from
the relativistic version of the potential model.

Although $S$ has no direct interactions with nuclei, it gets a magnetic
moment at one loop (with $\phi^\pm$ or $\mu^\pm$ in the loop),
depending upon $R \equiv m_S^2/m_\phi^2$
\be
\mu_S = {e|\lambda_2|^2 m_S \over 32\pi^2\,m_\phi^2 }f(R)
\ee
with
\bea	
	f(R) &=& \int_0^1 du\left( {u^2\over 1- Ru}
	+  {u(1-u)\over 1 - R + Ru}\right)
\eea
in the approximation of neglecting $m_\mu$ (see appendix \ref{DMDM}
for details).  The loop function
diverges logarithmically as $R\to 1$ in this approximation;
for example if $m_S = 0.9\, m_\phi$, $f(0.81) = 1.3$,
while for $m_S = 0.8\, m_\phi$, $f(0.64) = 0.93$.
	
If $\Sigma$ has spin, it inherits a magnetic moment from its $S$
constituents.  This will be the
case if $N_\hc$ is odd.  If $N_\hc=3$, the wave function
of $S^a S^b S^c$ is $\epsilon_{abc}$ in hypercolor space, and if
the spatial wave function is $s$-wave, total antisymmetry demands
that the spins be aligned, like the $\Delta$ baryon in QCD.  The quark
model then predicts that $\Sigma$ has a magnetic moment that is three
times that of $S$.  

The gyromagnetic ratio of $\Sigma$ is given (for odd $N_\hc$) by
$g_\Sigma = 4 N_\hc \mu_s M_\Sigma/e$.   Ref.\ \cite{Banks:2010eh} derived
an upper limit on $g$ versus DM mass based upon early XENON100 data
\cite{Aprile:2010um}.  We update their limit by comparing the relative
sensitivities of that search to the recent PandaX-II result
\cite{2017arXiv170806917P} in the high mass regime, to constrain
our model.  Allowing $m_S$ to vary between $\Lambda_\hc \cong
400\,$GeV (the value that maximizes $\zeta$ if $M=1\,$TeV) and 
1\,TeV, we rescale $\lambda_2^2$ by $(\zeta_{\rm max}/\zeta)^{1/2}$
where $\zeta_{\rm max} = 0.0037$ for the fiducial model parameters,
and $\zeta$ is smaller for $m_S < M$.  This keeps the predicted
$B$ decay signals constant while varying $m_S$, if all the couplings
scale in the same way.  
Using (\ref{msigma}), 
$g_\Sigma$  and $m_\Sigma$ are determined as a function of $m_S$.
The result for $N_\hc=3$ is shown in fig.\ \ref{msigma}(right).  
It translates to an upper limit $m_S \lesssim 800$\,GeV.  ($\zeta$
is not appreciably decreased at this value of $m_S$.)

%
\begin{figure}[t]
\hspace{-0.4cm}
\centerline{
\includegraphics[width=0.95\hsize]{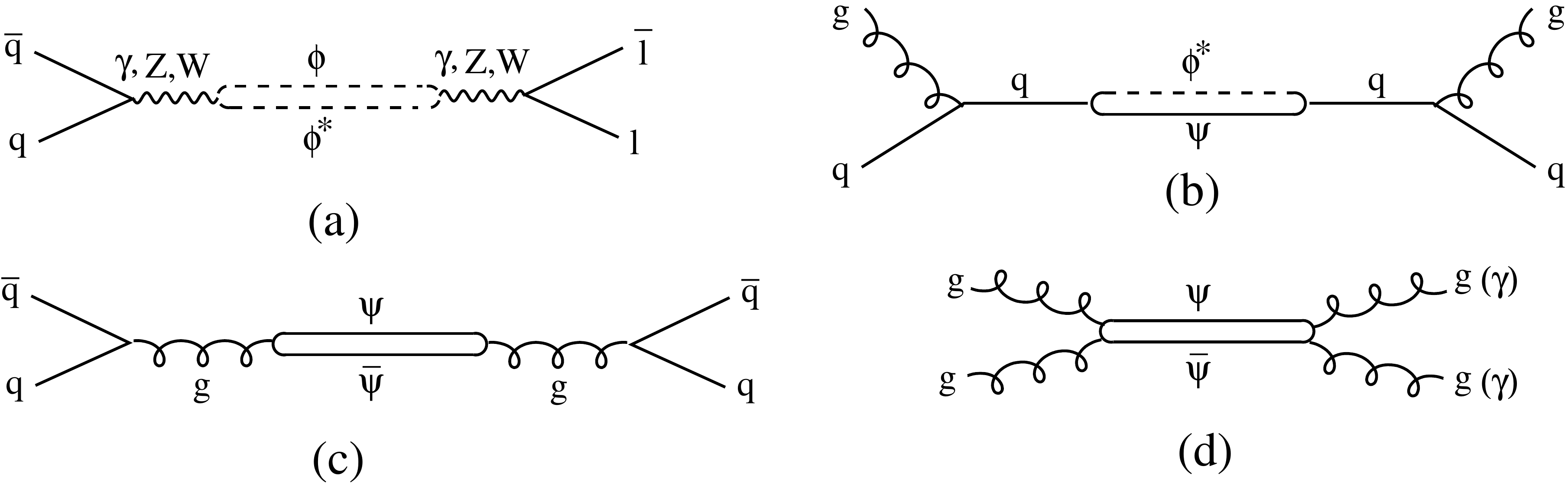}}
\caption{Resonant production of  HC bound states
leading to dileptons (a), dijets (b-d) or diphotons
(d).}
\label{lhc}
\end{figure}
%

\section{Collider constraints}
\label{collider}

Because of hypercolor confinement, only HC-neutral bound states of the
new particles $\phi$, $\Psi$, $S$ can be produced in high-energy
collisions.  We expect the dominant mode for discovery to be resonant
production of vector and scalar HC mesons, through the diagrams in
fig.\ \ref{lhc}.  We work in a region of parameter space where
$\Lambda_\hc$ is not too much lower than the constituent masses $M$,
which avoids the complications of quirks \cite{Kang:2008ea}.  In the
quirky regime, there are many excited quantum states to scatter into,
that can efficiently radiate hypergluons and settle to the ground
state before self-annihilating.  Then partons with energy above that
of the resonance could still be effective for production.   

On the other hand for $\Lambda_\hc\lesssim M$, the resonances are
well-separated and can only radiate QCD gluons or electroweak gauge
bosons since the HC glueball mass is too heavy, 
$\sim 7\Lambda_\hc$ \cite{Chen:2005mg} for $N_\hc = 3$.  In this regime, the parton center of  mass energy
must match the mass of the resonance, and we can use similar
techniques for computing resonant production of bound states as for
$J/\psi$ and $\Upsilon$ in $e^+e^-$ collisions.  This approximation 
misses events in which the bound state is produced by partons above
the threshold for the resonance, plus QCD radiation to take away the
excess energy.  We defer study of such higher-order corrections.

\subsection{Production cross section}

In general, the production cross section from colliding protons
to form a narrow resonance 
$R$ from partons $p_i$, $p_j$ can be expressed as 
\bea
\label{prod}
  \sigma(pp\to R) &=& {16\pi^2 \over m_R s} {(2J+1)N_c\, \Gamma(R\to p_i p_j)\over
(2s_1+1)(2s_2+1) N_{c1} N_{c2}}
	\\
	&\cdot&\int_{m_R/s}^1 {dx\over x} 
	\left[f_i(x) f_j(m_R/s x) + \{i\leftrightarrow j\}\right]\nn
\eea
where $f_i$ is the parton distribution (PDF) for $p_i$, $2s_i+1$ 
and $N_{ci}$ are the number of spin states and colors
respectively of the incoming particles, $2J+1$ and $N_c$ likewise for
the resonance, and $\Gamma$ is the partial width for the decay that is
inverse to the production process.
(In the case of incoming gluons, the doubling of PDFs in the second
line is correct, compensating for the factor of $1/2$ in the 
phase space for decay of $R$ into identical particles.)  The
generalization to resonances carrying SU(2)$_L$ rather than QCD
quantum numbers is obvious.

\subsection{Predicted widths}

The decay widths of bound states have been quantitatively  addressed
in regimes where the constituent masses are either much heavier than
the confinement scale \cite{Kang:2008ea} or much  lighter
\cite{Kilic:2008pm,Kilic:2009mi}.  We are interested in the case where
$\Lambda_\hc \sim 0.4\,M$ (with the possible exception of $m_S \ll
\Lambda_\hc$, but this is not relevant here because the $S\bar S$
bound states cannot be produced at LHC).  For $N_\hc=3$, the potential
model predicts that the kinetic energy of the constituent is $0.3$ of
its mass energy, so the nonrelativistic approximation is not very
bad.   Ultimately, lattice calculations should be done to make more
quantitative predictions.

We ignore decays of bound states into hypergluons, since we work in a
regime where any hyperhadrons  are too heavy to be produced.  Even if $m_S < \Lambda_\hc$, the
$\Pi_S = S\bar S$ pseudoscalar is not a light pseudo-Nambu
Goldstone boson, because the approximate chiral U(1) flavor symmetry
is anomalous, like for the $\eta'$ of QCD.  Likewise the glueballs
of the SU$(N_\hc)$ are also too heavy.

Ref.\ \cite{Kang:2008ea} calculates (in terms of $|\psi(0)|^2$) the
decay widths assuming that the bound state constituents do not form a
resonance of definite spin, but these are straightforward to rescale
for the physical eigenstates of spin.\footnote{For example, decay via
a single virtual gauge boson into fermions $f\bar f$ gets a factor of
$4/3$ to correct for the fact that only the spin-1 state contributes
in the average over spins of the initial constituents.  Similarly
decays to two gauge bosons occur only for spin-0 resonances, requiring
a factor of 4 correction to the spin-averaged rate. Likewise the
results of \cite{Kang:2008ea} ignore QCD color correlations of the 
constituents and must be rescaled for states of definite color.}\ \ 
Doing this gives results in agreement with the treatment of bound
state decays in ref.\ \cite{Peskin:1995ev}.  In our model, the
$\Psi\bar\Psi$ states can be in the QCD color singlet or octet
representations. Given that $\Lambda_\hc\gg \Lambda_{QCD}$, this distinction
is unimportant for the dynamics since the lifetime of the bound state
is much shorter than the QCD hadronization time scale.

\subsubsection{$\Psi\bar\Psi$ resonances}

Starting with the $\Psi\bar\Psi$ bound states, there is the
pseudoscalar $\Pi_\Psi$ that can be produced by gluon fusion
(diagram (d) of fig.\ \ref{lhc}) or the vector $\rho_\Psi$ coming
from $q\bar q$ annihilation, diagram (c). The respective decay widths are
\cite{Kang:2008ea}
\bea
\label{pp2gg}
	\Gamma(\Pi_\Psi\to gg) &=& {128\,\pi \over 27}
	N_\hc\,\alpha_s^2\,{|\psi(0)|^2\over m_{\Pi_\Psi}^2}\\
	\Gamma(\rho_\Psi\to u\bar u) &=& 
{8\,\pi \over 27}
	N_\hc\,\alpha_s^2\,{|\psi(0)|^2\over m_{\rho_\Psi}^2}\\
	\Gamma(\rho_\Psi\to e^+e^-) &=& {4\pi\over 9} N_\hc\,\alpha^2
\,{|\psi(0)|^2\over m_{\rho_\Psi}^2}
\eea
The last one, unlike the previous two, involves only the color singlet
state so it does not entail an average over QCD colors.  We include it
because it is potentially relevant for the dilepton final state.  
However counting
channels, it predicts a branching ratio of $B_{l^+l^-} =
(\alpha/\alpha_s)^2/2\cong 2.5\times 10^{-7}$ into electrons or muons,
implying that the dilepton channel is actually unimportant for this
resonance.

\subsubsection{$\phi\bar\phi$ resonance}

There are four kinds of $\rho_\phi = \phi\phi^*$ vector bound states,
three components in an SU(2)$_L$ triplet and one singlet.  Since the
SU(2)$_L$ dynamics are not important at the TeV scale, we average
over the isospins of the bound state constituents.  The angular
momentum of $\rho_\phi$ is purely orbital, so the constituents are in
a relative $p$-wave for which $\psi(0)=0$.  Therefore the width
depends upon the derivative of 
$\psi$ at the origin.  Following \cite{Cline:2013yya}, we find that
the SU(2)$_L$ contribution to the width is
\be
\Gamma(\rho_{\phi}\to W^*\to q_i\bar q_j) =
	{3\pi\over 2} 
	N_\hc\, \alpha_2^2\,{|\vec\nabla\psi(0)|^2\over 
	m_{\rho_\phi}^4}
\ee
for decay into a single generation of approximately massless 
quarks.  For decays into leptons, the result is $1/3$ times smaller
(from lack of QCD color), giving a branching ratio of $3/4$ to
quarks and $1/4$ to leptons. Hence this process 
leads to both dijet and dilepton signals.  For decays into muons,
there is an additional, potentially larger contribution from 
$S$ exchange (that interferes with the $W$ exchange contribution, but
we ignore this interference for purposes of estimation),
given by
\be
\Gamma(\rho_{\phi}\to \mu^+\mu^-) =
	{2\pi\over 3}	
	N_\hc\, \left(|\lambda_2|^2\over 4\pi\right)^2\,{|\vec\nabla\psi(0)|^2\over 
	(m_{\rho_\phi}^2 + \sfrac14 m_S^2)^2}
\ee

\subsubsection{Composite fermion}

For decay of the composite fermion to $qg$, taking $m_\phi\cong
m_\Psi$ as before, we find
\be
	\Gamma(F_q\to g q_i) \cong N_\hc 
	{\alpha_s|\tilde\lambda_i^{(\prime)}|^2\over 2 m_\Psi^2}|\psi(0)|^2
\ee
where $i$ is the generation index of the quark and as usual the prime
is for couplings to up-type quarks.  The squared coupling must be
absorbed into sum over quark flavors in the parton luminosity factor
for production via $qg$ fusion.

\subsubsection{Vertex correction}
For all of the annihilation decays, a more quantitative estimate can
be made by taking account of perturbative corrections that dress
the annihilation vertex.  For charmonium and upsilon, the correction
from a gluon loop is known to be 
important, and we adopt the analogous correction for the HC
gluon exchange, with $\alpha_\hc$ evaluated at the scale of the
constituent mass.  The correction factors depend upon the annihilation
process, and in analogy with QCD we take 
\cite{Kwong:1987ak}
\be
	C = 1 -{(N^2-1)\alpha_\hc\over 8\pi}\left\{\begin{array}{cc}
	(20-\pi^2)/3,& \Pi_\psi\to gg\\	  
	 {16/3},& \rho_\psi \to f\bar f\end{array}\right\}
\label{corr_fact}
\ee
with $\alpha_\hc$ evaluated at the scale $\mu_*$, twice the inverse
Bohr radius.  We have assumed
that the known $N_\hc=3$ results generalize to other values by
rescaling by the number of hypergluons.

%
\begin{figure}[t]
\hspace{-0.4cm}
\centerline{
\includegraphics[width=0.95\hsize]{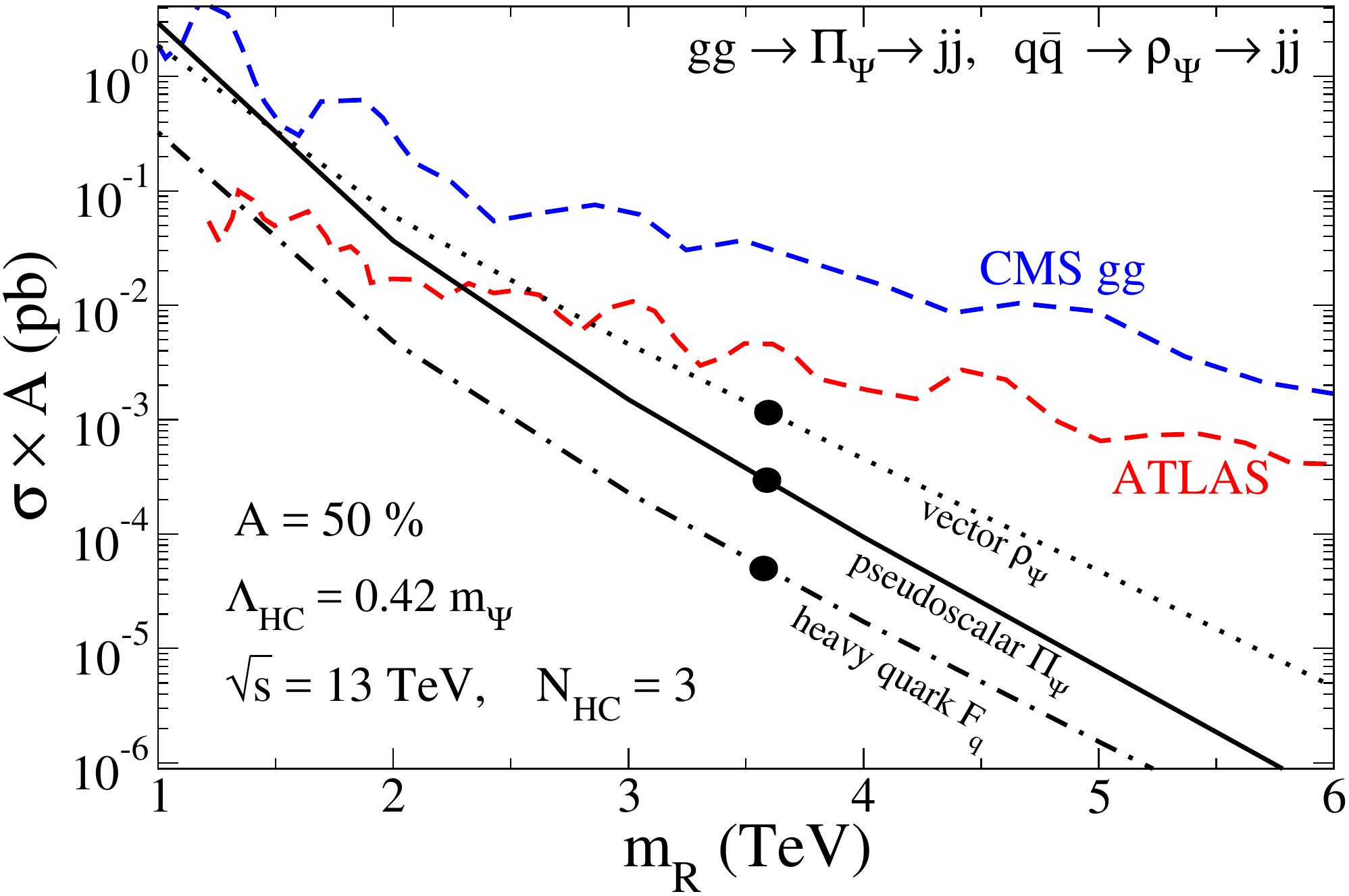}}
\caption{Dashed curves: limits on resonant dijet production 
cross section times
acceptance  versus resonance mass from ATLAS \cite{Aaboud:2017yvp} and CMS
\cite{CMS:2017xrr}, along with predictions for $\Pi_\Psi$ 
pseudoscalar (solid curve), $\rho_\psi$ vector (dotted curve)
and heavy quark $F_q$ (dot-dashed curve)
bound states, in the case of $N_\hc = 3$, acceptance
$A = 100\%$ and $\sqrt{s}=13\,$TeV center of mass energy.  
Predictions for the optimal value (in terms of LHC sensitivity) $\Lambda_\hc/m_{\Psi} = 0.4$ are
shown.}
\label{dijet}
\end{figure}
%

\subsection{Resonant search constraints}
Combining the decay widths with the general formula for the production
cross section (\ref{prod}), we find that it depends upon the same
combination $\zeta$, eq.\ (\ref{zeq}) that appeared in the dimension-6
FCNC operators.
One finds a 
production cross section for the color octet plus singlet states
$\Pi_\Psi$ given by
\be
	\sigma(pp\to \Pi_\Psi) = {8\,\pi^3\alpha_s^2 N_\hc\over  3 s}
        {\zeta C}{\cal L}_{gg}
\label{pppsi}
\ee
where ${\cal L}_{gg}$ is the parton luminosity factor for 
gluon fusion.  The perturbative correction (\ref{corr_fact}) is $C = 0.5$ at $m_\Psi = 
2.5\Lambda_\hc$, similar to the value that occurs in QCD 
for $J/\psi$, and $\zeta = 0.0037$ as in the previous sections. 
In fig.\ \ref{dijet}
we plot this prediction along with current constraints from ATLAS
\cite{Aaboud:2017yvp}
and CMS \cite{CMS:2017xrr} resonant dijet searches, assuming an 
acceptance of events passing experimental cuts of $A= 50\%$,
comparable to that of various models 
tested in the searches.  The resonance mass must exceed 2.3\,TeV,
which is satisfied for our fiducial model with $m_\Pi = 3.6\,$TeV.
Keeping $m_\Psi/\Lambda_\hc$ fixed, this limit would allow $m_\Psi$
to be no lower than 650 GeV. 

Similarly for production of the color octet vector $\rho_\Psi$ resonance for
$N_\hc=3$ we find
\be
	\sigma(pp\to \rho_\Psi) = {64\,\pi^3\alpha_s^2 N_\hc\over 9\,s}
        {\zeta C}{\cal L}_{q\bar q}
\ee
The loop correction factor (\ref{corr_fact}) is 0.7 for this
process.
The corresponding prediction is also plotted in fig.\ \ref{dijet}.\footnote{It is
possible that the true limit from the vector resonance is weaker 
since we have assumed 100\%
branching ratio into two jets.  In the QCD charmonium and upsilon systems,
decay to three gluons is much more likely, resulting in three or more
jets.  In this case the reconstruction of the resonance is 
challenging, and has not been carried out in experimental searches, to
our knowledge.  However the three-jet process is higher order in
$\alpha_s$ and should therefore be subdominant at the TeV scale.}\ \ 
We infer a limit of $m_\Psi > 820\,$GeV for the fiducial model.  
At larger values
of $m_\Psi/\Lambda_\hc = 5,\,10$, this limit becomes more stringent,
$m_\Psi\gtrsim 1\,$TeV, approximately independent of the value of
$m_\Psi/\Lambda_\hc$.

%
\begin{figure}[t]
\hspace{-0.4cm}
\centerline{
\includegraphics[width=0.95\hsize]{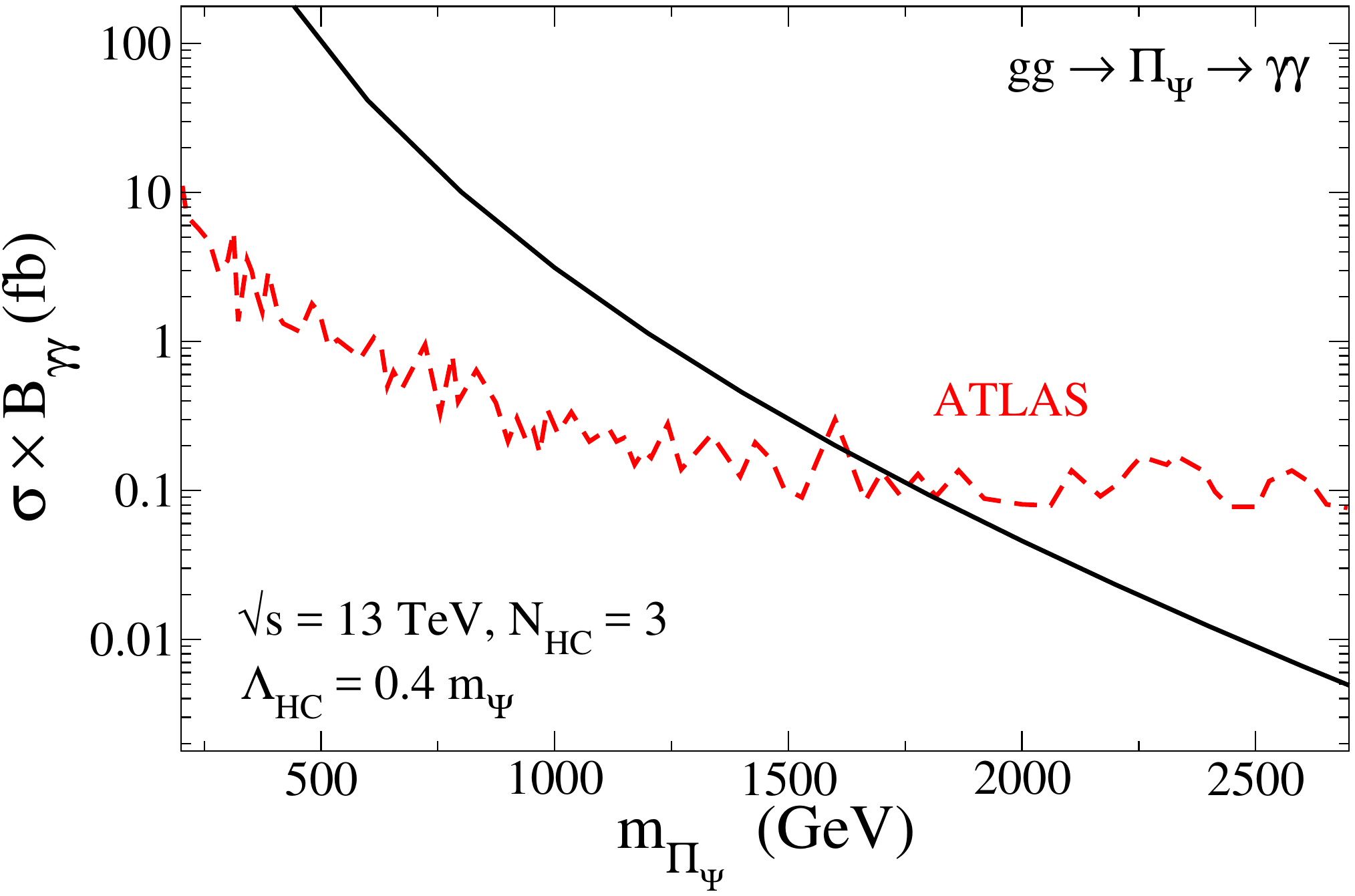}}
\caption{Production cross section for pseudoscalar bound state
$\Pi_\Psi$ times branching ratio for decay to two photons.  Dashed
curve is ATLAS limit \cite{Aaboud:2017yyg}, solid curve is
 predicted value for $N_\hc =
3$ and $\Lambda_\hc/m_\Psi =0.4$.}
\label{diphoton}
\end{figure}
%
%
\begin{figure}[t]
\hspace{-0.4cm}
\centerline{
\includegraphics[width=0.95\hsize]{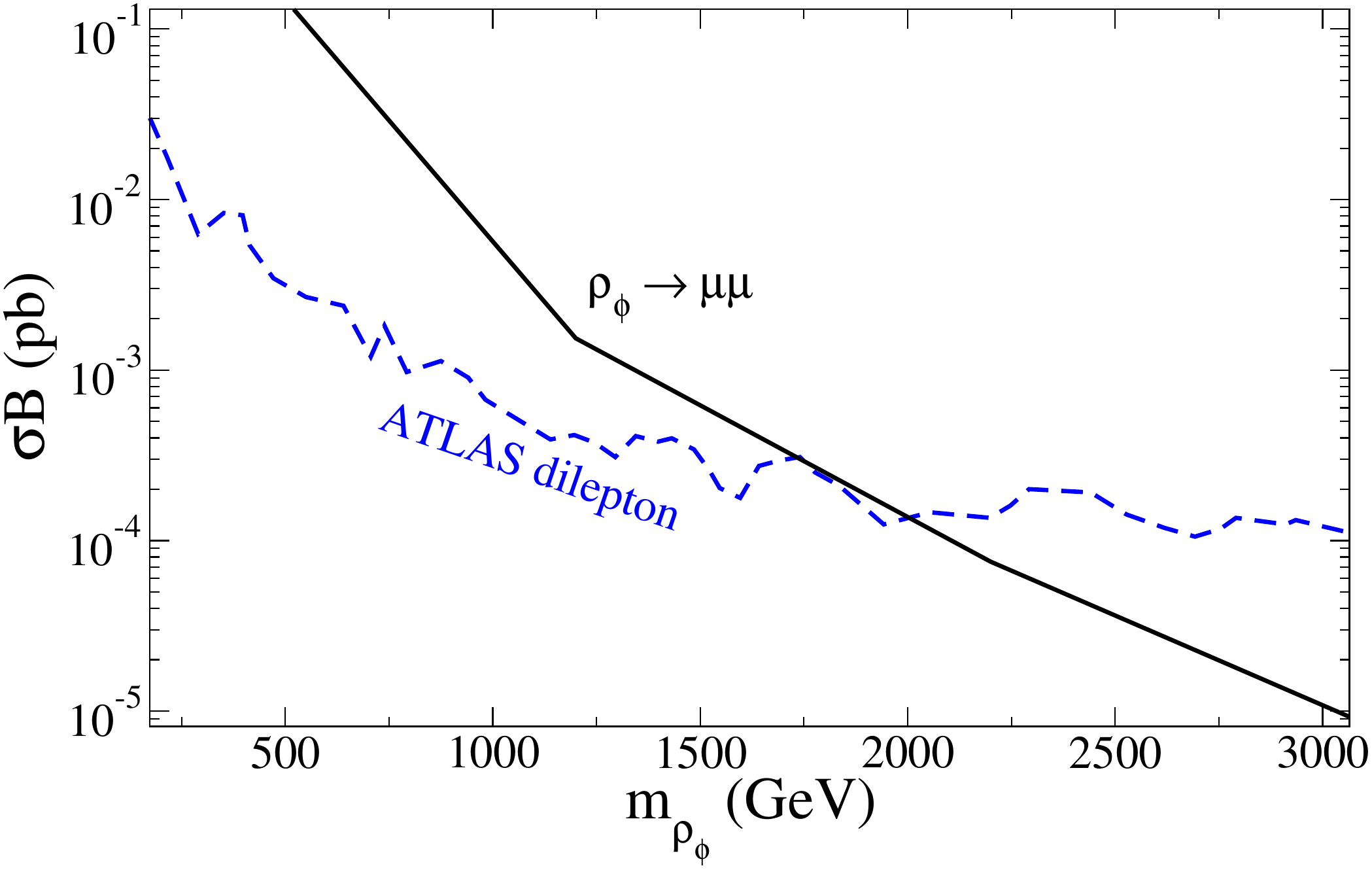}}
\caption{Production cross section for pseudoscalar bound state
$\rho_\phi$ times branching ratio for decay to two leptons.  Dashed
curve is ATLAS limit \cite{Aaboud:2017buh}, solid curve is
 predicted value for $N_\hc =
3$ and $\Lambda_\hc/m_\Psi =0.4$.}
\label{dilepton}
\end{figure}
%

In addition to the $\Pi_\Psi\to gg$ channel, the color singlet pseudoscalar can
decay to two photons.  Its decay width into gluons is smaller than
the color average (\ref{pp2gg}) by a factor of 16/9, and its 
branching ratio into photons is $B_{\gamma\gamma} = 9\alpha^2
q_\psi^4/2\alpha_s^2$  \cite{Craig:2015lra}.  The cross section for 
$pp\to\Pi_\Psi\to \gamma\gamma$ is therefore 
$(9/128)B_{\gamma\gamma}$ times eq.\
(\ref{pppsi}).  The predicted cross section is shown along with the
ATLAS diphoton limit \cite{Aaboud:2017yyg} in fig.\ \ref{diphoton};
it constrains $m_\Pi > 1.7\,$TeV.

The $\rho_\phi$ resonance has electroweak production cross section 
\be
	\sigma(pp\to \rho_\phi) = {2\,\pi^3\alpha_2^2 N_\hc\over
	 s}\,\zeta_p\,
        {\cal L}_{q\bar q}
\ee
where $\zeta_p =  |\nabla\psi(0)|^2/ m_{\rho_\phi}^5 = 
(\mu_*/m_{\rho_\phi})^5/96\pi \cong 5\times 10^{-5}$ from the 
potential model for $p$-wave states, evaluated at $\Lambda_\hc =
0.4\,m_\phi$.  The ATLAS dilepton limit 
\cite{Aaboud:2017buh}, slightly rescaled to account for dominant
decay into of $\rho_\phi$ to muons  
(see ref.\ \cite{Cline:2017ihf}  for details)
is plotted in fig.\ \ref{dilepton} along with the prediction for
branching ratio of 100\% into $\mu^+\mu^-$ (this neglects decays into
electroweak gauge bosons).  The resulting limit $m_{\rho_\phi} >
2\,$TeV is stronger than that from diphotons but weaker than that from
dijets for the $\Pi_\Psi$ and $\rho_\Psi$ states respectively.

The cross section for heavy quark $F_q$ production is
\be
	\sigma(pp\to F_q) = {N_\hc \alpha_s\over 2 s}\,\zeta\,
	\sum_i |\tilde\lambda_i^{(\prime)}|^2 {\cal L}_{g q_i}
\label{Fqprod}
\ee
The result assuming couplings (\ref{params}) and 100\% branching into
dijets is plotted in fig.\ \ref{dijet}, giving a weaker constraint
than those from $\rho_\Psi$ and $\Pi_\Psi$ resonant production.

%
\begin{figure}[t]
\hspace{-0.4cm}
\centerline{
\includegraphics[width=0.95\hsize]{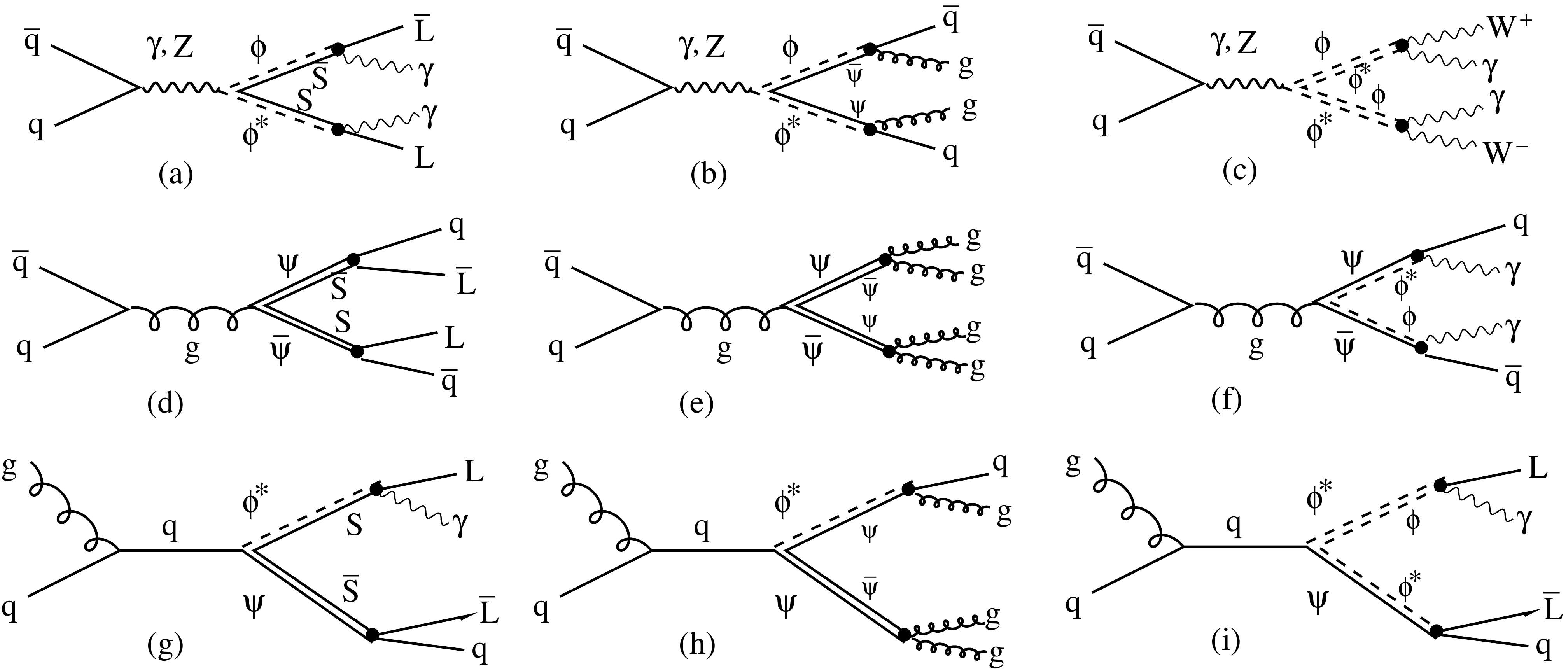}}
\caption{Pair production of composite states}
\label{lhc3}
\end{figure}
%

\subsection{Pair production of bound states}
The wave function at the origin (the $\zeta$ factor) suppresses
the predicted resonant production signficantly below the cross section
that would govern the creation of free pairs of hypercolored 
constituents.  This penalty is avoided if bound states are
pair-produced by hadronization in the HC sector, as depicted in 
fig.\ \ref{lhc3}.  Then each possible kind of final state 
will occur with a relative probability 
not much smaller than $1$ since for every pair there are only three kinds of particles
($\Psi$, $\phi$, $S$) that can appear from the vacuum.  The only 
other
bottleneck is that the incoming partons must have at least enough energy
to make both bound states.  Then the cross section averaged over
PDFs takes the form
\be
	\sigma = \int_0^1 dx_1 \int_0^1 dx_2\, f_1(x_1) f_2(x_2) 
	\Theta(x_1 x_2 s - 4 m_c^2)\,\hat\sigma
\ee
where $m_c$ is the mass of the composite state.
The largest cross sections are for $q\bar q\to \Psi\bar\Psi$ and
$gg\to\Psi\bar\Psi$,
\bea
	\hat\sigma_{q\bar q} &=&{ 8\pi\alpha_s^2\,N_\hc\over 27\, \hat
s^2}
	\left(2m_\Psi^2 + \hat s\right)\chi\\
	\hat\sigma_{gg} &=& {\pi\alpha_s^2\,N_\hc\over 3\, \hat s}
	\Biggl(-{\chi\over 4}\left(7 + 31 m_\Psi^2/s\right)\nn\\
	&+& \left(1 + 4 m_\Psi^2/\hat s + m_\Psi^4/\hat s^2\right)
	\ln\left(1+\chi\over 1-\chi\right)\Bigg)
\label{ggpsipsi}
\eea
with $\chi = \sqrt{1 - 4 m_\Psi^2/\hat s}$ and $\hat s = x_1 x_2 s$.

We find that the composite mass $m_c$ must be as small as possible to
get interesting  constraints from this process.  This occurs for
$m_S\ll\Lambda_\hc$ in the $\Psi\bar S$ bound states, for which the
potential model gives
\be
	m_c \cong (0.75 + 1.8 N_\hc)\Lambda_\hc + m_\Psi
\ee
These are leptoquark states that decay into $q\ell$, and have been
searched for by ATLAS \cite{Aaboud:2016qeg} and CMS 
\cite{CMS:2016qhm,Sirunyan:2017yrk}.  The searches look for two jets
and two leptons, either $\mu^+\mu^-$ or $\tau^+\tau^-$ 
respectively, for second or third generation leptoquarks.  The limits
are on the production cross section times $\beta^2$ where $\beta$
is the branching ratio into the final state searched for.  In our
model, $\beta \le 1/2$ since $\rho$ can decay into either a charged
lepton or neutrino with equal probability.  Moreover the branching
ratio into muons is $|\lambda_2|^2/(|\lambda_2|^2 + |\lambda_e|^2$,
and that into tau is $|\lambda_3|^2/(|\lambda_2|^2 + |\lambda_e|^2$,
typically giving a further reduction in $\beta$.  (Recall that LFV
constraints on $\lambda_1$ make the branching into $e^+e^-$ negligible
in our model.)

%
\begin{figure}[t]
\hspace{-0.4cm}
\centerline{
\includegraphics[width=0.95\hsize]{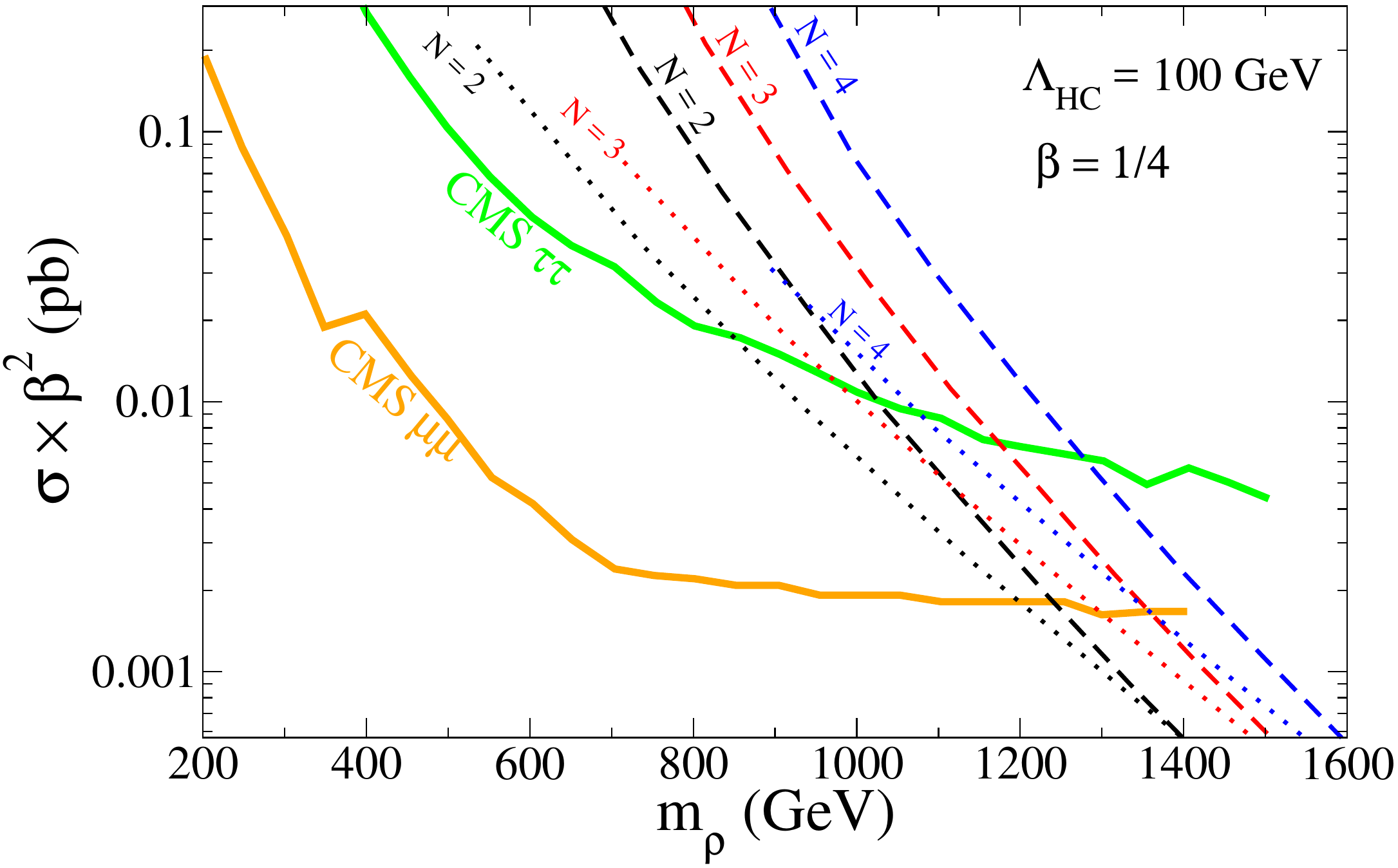}}
\caption{Cross section for pair production of composite leptoquarks
times branching ratio squared for decays into $\mu$ or $\tau$ plus
jet ($j$).  Solid is CMS upper limit for $jj\mu\mu$ 
\cite{CMS:2016qhm} or 
$jj\tau\tau$ \cite{Sirunyan:2017yrk} final states, dotted is prediction from $q\bar q\to
\rho\rho^*$, dashed is prediction from $gg\to \rho\rho^*$, assuming 
low compositeness scale $\Lambda_\hc = 100\,$GeV and massless 
$S$ constituents.}
\label{lq-pair}
\end{figure}
%

The predicted cross sections and observed limits are shown in fig.\
\ref{lq-pair} for a model with $\Lambda_\hc = 100\,$GeV and 
$m_S\ll\Lambda_\hc$, which is a limiting case for lowering the
masses of bound states containing $S$.  We have assumed
$\lambda_3\sim\lambda_2$ so that $\beta \cong 1/4$.  The most
stringent limits come from the search for final state muons, with 
production via the $gg\to\Psi\bar\Psi$ cross section
(\ref{ggpsipsi}).  It requires the leptoquark mass to be $\gtrsim
1.1,\,1.2,\,1.26\,$TeV for $N_\hc = 2,3,4$, hence $m_\Psi \gtrsim 
670,\,590,\,470\,$GeV respectively.

\begin{figure}[t]
\hspace{-0.4cm}
\centerline{
\includegraphics[width=0.95\hsize]{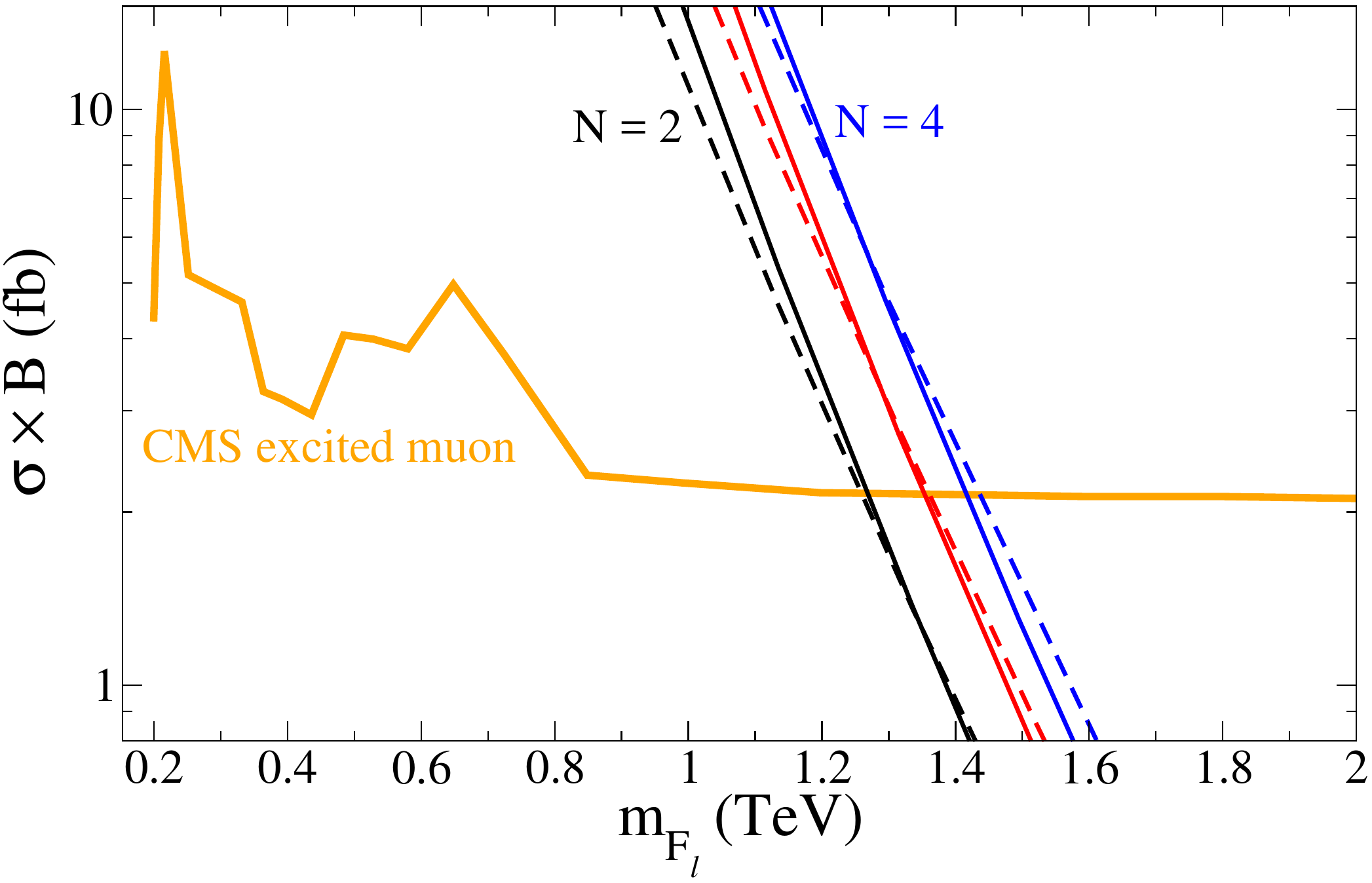}}
\caption{Cross section for  production of composite heavy 
leptons decaying to  $\mu$  plus
photon.  Solid (orange) curve is CMS upper limit, dashed are predicted
values from $F_l$ pair production, assuming 
low compositeness scale $\Lambda_\hc = 100\,$GeV and massless 
$S$ constituents.  Neighboring solid curves are from production
of $F_l$-leptoquark pairs, eq.\ (\ref{fllq}).}
\label{excited-lepton}
\end{figure}

Other bound states not involving $S$ as a constituent, although they
have potentially interesting signals, are too heavy to be produced
at a significant level relative to current constraints from LHC.
The composite lepton $F_l = S\phi^*$ arises from electroweak
$q\bar q\to \phi\phi^*$ pair production, with parton-level
cross section
\be
	\hat\sigma = {\pi\alpha_2^2 N_\hc \over 4 }{\hat s\left(1 - 4
	M^2/\hat s\right)^{3/2}\over (\hat s-m_W^2)^2 + m_W^2\Gamma_W^2}
\ee
in the simplifying approximation $g'\ll g$ ($\sin\theta_W\to 0$).
CMS has searched for excited muons and taus decaying to the normal
state plus photon \cite{CMS:2016sgu}.  Figure \ref{excited-lepton}
shows the limit on excited $\mu$ production, which is more stringent
than that for excited $\tau$, versus the predicted cross sections (100\% branching
of $F_l\to \mu\gamma$ is assumed).  Again we take the extreme choice
with $\Lambda_\hc = 100\,$MeV and $m_S = 0$ to obtain light enough
$m_{F_l}$ to fall within the currently probed mass range; larger
values of $\Lambda_\hc$ and $m_S$ are not excluded.   Comparison
with fig.\ \ref{lq-pair} shows that $F_l$ is excluded up to similar masses as
the leptoquarks, $m_{F_l}\lesssim 1.3-1.5\,$TeV, depending upon 
$N_\hc$.

The remaining mechanism to produce the lightest states is diagram 
(g) of fig.\ \ref{lhc3}, which has partonic cross section
\be
	\hat\sigma = N_\hc{\alpha_s |\tilde\lambda_i^{(\prime)}|^2\over
24\, \hat s}
	\sqrt{1-4M^2/\hat s}
\label{fllq}
\ee
neglecting the quark mass (a good approximation since the top
quark PDF is very small).  This produces the heavy lepton $F_l$
in conjunction with a leptoquark, and would be probed by the CMS
search \cite{CMS:2016sgu}.  The predicted cross section (assuming
100\% $F_l\to\mu\gamma$) is also shown in fig.\ \ref{excited-lepton} and happens
to be close to that coming from $F_l$ pair production.  Since the
leptoquark is produced singly, it does not produce the same signal
as for the existing searches that assume pair production and would 
require a new analysis, searching for the unusual final state of
two leptons, a jet and a photon.

%
%

\section{Summary and Conclusions}

We have presented a particularly simple realization of composite
leptoquarks to explain the anomalous $B$ decay ratios $R_{K^{(*)}}$, as
well as providing a composite dark matter candidate.  The new
ingredients are an SU($N_\hc$) confining gauge sector under which three
types of matter fields $\Psi$, $S$, $\phi$ transform as fundamentals. 
Only two kinds of interactions with SM fermions are allowed, and these
break the approximate flavor symmetries of the SM through the six
couplings $\lambda_i$ to leptons and $\tilde\lambda_i$ to quarks.
The leptoquark is the $\Psi\bar S$ vector bound state.

There is a limited region of parameter space allowed by FCNC
constraints on meson-antimeson oscillations (mediated by the
$\Psi\bar\Psi$ vector bound states), direct dark matter
searches, and LHC searches for resonant dijets, which require the
$\Psi$ and $\phi$ constituents to be near the TeV scale, and the
new confinement scale $\Lambda_\hc$ must be $\gtrsim 100$\,GeV.  
(Our benchmark model takes $\Lambda_\hc=400$\,GeV.)\ \ The dark
matter is a baryon-like bound state $\Sigma = S^{N_\hc}$ with mass 
$\sim(1-5)\,$TeV.  If $N_\hc$ is odd, $\Sigma$ can have a magnetic dipole
moment that requires $m_\Sigma < 2.5\,$TeV to avoid direct detection.

For very light dark matter constituent and low confinement scale  with
$m_S\ll\Lambda_\hc\sim 100$, the new bound states (leptoquarks and
heavy lepton partners) are light enough ($\sim 1\,$TeV) to be
pair-produced and significantly constrained by ATLAS or CMS  searches
for leptoquarks and excited leptons.  The model also predicts
production of the leptoquark and heavy lepton in association, leading
to the unusual signal of two muons, a jet and a photon. The dark
matter dipole moment is below direct detection constraints for small
$m_S$.

The predictions are sensitive to nonperturbative aspects of
confinement, especially bound state masses and wave functions. We have
used a simple potential model that works reasonably well for known
examples from QCD, but which is not necessarily reliable for the
parameters of interest here, and neglects spin interactions between
the fermionic consituents.  Lattice studies of the masses and decay
constants would allow for a useful improvement.  Nevertheless it seems
likely that the kind of model presented here should give rise to 
other anomalous signals in FCNC, dark matter or collider searches.
It is encouraging that the experimental status of the 
$B$ decay anomalies should become more clear from future data at LHCb
and Belle II during the next few years \cite{Albrecht:2017odf}.

\bigskip
{\bf Acknowledgment.}  I thank I.\ Brivio, J.\ Evans, B.\ Gavela, J.\ Jiang, 
D.\ London, J.\ Martin Camalich, Y.\ Kats, M.\ Redi, M.\ Trott and  for helpful discussions,
and A.\ Urbano for pointing out ref.\ \cite{Guberina:1980dc}.  This
work was supported by NSERC (Natural Sciences and Engineering Research
Council of Canada).

\appendix

\section{Potential model for bound states}
\label{app3}

\subsection{Nonrelativistic constituents}

We first consider bound states where all the constituents are 
heavier than $\Lambda_\hc$ and hence nonrelativistic.  
Following \cite{Cline:2013yya,Alves:2010dd}, we estimate the size of
hypercolor mesonic bound states using a variational method, with hydrogen-like
ans\"atze for the wave functions.  The Coulomb-like contribution to
the potential from hypergluon exchange is \cite{Raby:1979my}
\be
	V_c = -{\alpha_\hc\over 2 r}\left(N_\hc - {1\over N_\hc}\right)
	\equiv -c_\alpha{\alpha_\hc\over r}
\ee
for a hyperquark-antiquark pair in the hypercolor singlet state. 
There is a linear confining potential 
\be
	V_l = \sigma r
\ee
where $\sigma \cong 2(N_\hc-1)\Lambda_\hc^2$ is the string tension
and $\Lambda_\hc$ is identified with the $\Lambda$ parameter of the 0-flavor
running coupling in the $\overline{\rm MS}$ scheme.
The coefficient is deduced for the case of $N_\hc=3$ from lattice
studies \cite{Teper:1998kw} and its generalization to other values of
$N_\hc$ is inferred from large-$N$ scaling.  The energy of an $s$-wave
state with wave function 
\be
	\psi_s(r) = {\mu_*^{3/2}\over \sqrt{8\pi}} e^{-\mu_*r/2}
\ee
is
\be
	E_s = m_1 + m_2 + {\mu_*^2\over 8 m_r} - { c_\alpha\over 2}\alpha_\hc\mu_* + 
	3{\sigma\over\mu_*}
\label{Eswave}
\ee
where $m_r$ is the reduced mass of the 
constituent particles, and $\alpha_\hc$ implicitly depends on 
$\mu_*$ since it is the
running coupling evaluated at that scale.   We thus minimize
the energy numerically.
We use the 4-loop, 0-flavor expression for the running coupling given
in \cite{PDG2,Czakon:2004bu}.  For $\mu_*<\Lambda_\hc$ the perturbative
coupling diverges and we cut it off at $\alpha_\hc = 2$.  (Larger
cutoffs lead to numerical artifacts in the minimization.)

This model gives reasonable results for $J/\psi$ and $\Upsilon$
using $m_c = 1.3\,$GeV, $m_b = 4.2\,$GeV for the quark masses, and
$\Lambda_{\rm\sss QCD} = 200\,$MeV.
For these systems $\psi(0)$ is measurable through the electromagnetic
decays via $\gamma^*$ to $e^+e^-$.  For $J/\psi$ we find
$\psi(0)^2 = 1/(0.5\,{\rm fm})^3$, close to the measured value
$1/(0.53\,{\rm fm})^3$, while the mass is predicted to be $3.0$\,GeV,
compared to the measured $3.1$\,GeV.  The model predicts 
$\psi(0)_{\Upsilon}^2/\psi(0)_{J/\psi}^2 = 4.6$, while experimentally
one finds 4.7, by
comparing the decay rate formula (including the correction
(\ref{corr_fact})) to the observed partial widths.  The worst
agreement is for $m_\Upsilon$, predicted to be 8.6\,GeV versus the
measured value 9.5\,GeV.

For $p$-wave states with 
\be
	\psi_p(r) = {\mu_*^{5/2}\over \sqrt{96\pi}}\, r e^{-\mu_* r/2}
\ee
we find the energy
\be
	E_p = m_1 + m_2 + {\mu_*^2\over 8 m_r} - { c_\alpha\over 4}\alpha_\hc\mu_* + 
	5{\sigma\over\mu_*}
\label{Epwave}
\ee

For baryons in an $s$-wave state, we take the ansatz $\psi \sim
e^{-\mu_*(r_1 + r_2 + \cdots)}$ and the energy is
\bea
	E_b &=& \sum_{i=1}^{N_\hc}\left( m_i  + {\mu_*^2\over 8\,m_i}
\right)
	\\
	 &+& \sfrac12 N_\hc (N_\hc-1)\left(-{5 c_\alpha\,\alpha_\hc\mu_*\over
16(N_\hc-1)} + 
	f{35\sigma\over 8\mu_*}\right)\nn
\eea
(note that the Coulomb attraction between $qq$ in an antisymmetric
hypercolor state is $1/(N_\hc-1)$ times weaker than that 
between $q\bar q$).
Comparison to the nucleon of QCD motivates the correction factor $f = 0.065$,
indicating that the string tension is much smaller between $qq$ than
$q\bar q$ states.

\subsection{Heavy-light or relativistic systems}

In the case where $m_S < \Lambda_\hc$, one or more constituents 
is relativistic.  Then in the center of mass
system of a heavy-light meson we have mass plus kinetic energy
$m_1 + p^2/(2 m_1) + \sqrt{p^2 + m_2}$ in the Hamiltonian
\cite{Rai:2002hi}.  Still using the wave function ansatz
$\psi \sim e^{-\mu_* r/2}$ for $s$-wave states, the expectation value
of $\sqrt{p^2 + m^2}$ can be found by Fourier transforming $\psi$
and evaluting the integral in the momentum eigenstate basis.  The
result is a complicated analytic function that can be fit to the
simpler form 
\be
	\left\langle\sqrt{p^2+m^2}\right\rangle\cong
	\mu_* \left(0.84 + 0.52\left(m\over
\mu_*\right)^{1.6}\right)
\ee
(At $m=0$ the exact result is $8\mu_*/(3\pi)$.)
We find that this approximation is good to better than $1\%$
for $m<\mu_*$.  
The kinetic energy in (\ref{Eswave},\ref{Epwave}) is then replaced
by
\be
	m_1 + {\mu_*^2\over 8 m_1} + 
	\left\langle\sqrt{p^2+m_2^2}\right\rangle
\ee
For baryons with some constituents light, the appropriate replacement
is obvious.  The coupling $\alpha_\hc$ should be evaluated taking into
account running with the appropriate number of light constituents.

\section{Transition moments from compositeness}
\label{app2}
Here we semiquantitatively estimate the transition magnetic moment
interactions between heavy composite fermions $F_\R$ and their 
elementary SM counterparts, $f_\R$, following the same formalism 
used by ref.\ \cite{Guberina:1980dc} for radiative decays of the 
$Z$ boson to quarkonium bound states.  The formalism is appropriate
for nonrelativistic systems, which is not a very good approximation 
in our case since we prefer the constituent masses and confinement
scale to be of the same order.  Hopefully it gives a reasonable
estimate, which would require a dedicated lattice study to improve
upon.

The amplitude for $F\to
f_i\gamma$ can be written as
\be
	A(F\to f_i\gamma) = 
-i\overset{\scriptscriptstyle(\sim)}{\lambda_i}\bar u_{f_i}\int{d^{\,4}q\over (2\pi)^4}\,
	{\cal O}_\gamma(q,k,Q)\, \chi(Q,k)
\ee
where the momenta are indicated in fig.\ \ref{bound-state}.  $Q$ is the momentum
of they decaying heavy $F$, $k$ is that of the photon, 
$q$ is the relative
momentum between the heavy fermion constituents and 
$\chi(Q,q)$ is the Bethe-Salpeter wave function for the bound
state.  The operator ${\cal O}_\gamma$ comes from the part of the diagram
that is introduced by insertion of the photon vertex, which depends
upon what kind of particle $F(f)$ is.  For down-like quarks, the photon
attaches to both $\Psi$ and $\phi^\pm$; for up-like quarks it attaches
only to $\Psi$ (since $\phi^0$ is the other constituent), while for
leptons it attaches only to $\phi^\pm$ since $S$ is neutral.

For simplicity we  take the constituent masses to be equal so that
in the absence of exchanged momenta, each carries half of the momentum
$Q^\mu$ of the bound state.  The
general form of ${\cal O}$ can then be written as 
\bea
	{\cal O}_\gamma &=& -ie q_\phi{i(-Q + 2q + k)^\mu\over (-Q/2+q + k)^2-
	m_\phi^2}P_\R\nn\\
	&-& ie q_{\psi
(S)}{i\over\slashed{Q}/2+\slashed{q}-\slashed{k}-m_{\psi(S)}}
	\gamma^\mu P_\R
\eea
where $P_\R$ projects onto right-handed chirality.  The spinor wave
function takes the form
\be
	\chi(Q,q) = {2\pi \over \sqrt{2\mu}}\, \delta\left(q_0 -{\vec
q^{\,2}\over 2\mu}\right)u_{\psi(S)}(Q/2+q)\,\tilde\psi(q) 
\ee
where $\mu$ is the reduced mass of the bound state constituents, 
$u_{\psi(S)}$ is the Dirac spinor for the fermionic member, and 
$\tilde\psi$ is the Fourier transform of the spatial wave function.

For nonrelativistic systems, the wave function is strongly peaked at
small $\vec q$ and it is a good approximation to set $q=0$ in ${\cal
O}_\gamma$. Then the factor $\int d^3q/(2\pi)^3  \tilde\psi(q) = 
\psi(0)$, the spatial wave function evaluated at the origin, and the
amplitude becomes
\be
	A(F\to f_i\gamma) = 
	-i{\overset{\scriptscriptstyle(\sim)}{\lambda_i}\psi(0)
\over \sqrt{2\mu}}\,\bar u_{f_{i}}
	{\cal O}_\gamma(0,k,Q)\, P_\R\ u_{\psi(S)}
\ee
It is pertinent to compare this to the corresponding amplitude with no
photon, which is just the mass mixing amplitude,
\be
	A(F\to f_i) = 
	-i{\overset{\scriptscriptstyle(\sim)}{\lambda_i}\psi(0)
\over \sqrt{2\mu}}\,\bar u_{f_{i}}
	\, P_\R\, u_{\psi(S)} = -i\mu^i_f \,\bar u_{f_i}
	\, P_\R\, u_{\psi(S)}
\ee
This allows us to infer that the transition moment is related to the
mass mixing in a definite way.

For leptons, the decay amplitude from diagram (a) of fig.\ 
\ref{bound-state} is
\bea
	A(F_\ell\to \ell_i\gamma) &=& -ie \mu^i_\ell
	\,\bar u_{\ell_{i\L}}
	{(Q-k)^\mu\over \sfrac14 m_{F_\ell}^2 + 
	m_\phi^2}  \,u_{S_\R}\nn\\
	&\to& -ie{\mu^i_\ell\over m_{F_\ell}^2} \bar u_{\ell_{i\L}}i\sigma^{\mu\nu}
	k_\nu \, u_{S_\R}
\eea
where in the second line we used the Gordon identity 
\be
(p+p')^\mu \bar u'_\L u_\R = m'\,\bar u'_\R\gamma^\mu u_\R + 
 m\,\bar u'_\L\gamma^\mu u_\L + \bar u'_\L i\sigma^{\mu\nu}
	k_\nu u_\R
\ee
and took
$m_\phi\simeq m_{F_\ell}/2$.  The extra terms going as $\gamma^\mu$ in
the Gordon identity are canceled by diagrams (c) and (d) of fig.\ 
\ref{bound-state}, which serves as a check on the sign.  We ignored 
terms going as $k^\mu$ that do not contribute to the amplitude because
of transversality of the photon polarization vector.

Similarly for up- and down-type quarks respectively we obtain 
\bea
	A(F_u\to u_i\gamma) &=& 
	 i{2e\over 3}{{{\mu'}^i_u}\over m_{F_u}^2} \bar u_{u_{i\L}}
	i\sigma^{\mu\nu}
	k_\nu \, u_{\Psi_\R}\nn\\
	A(F_d\to d_i\gamma) &=& 
	 -i{e\over 3}{{{\mu}^i_d}\over m_{F_d}^2} \bar u_{d_{i\L}}
	i\sigma^{\mu\nu}
	k_\nu \, u_{\Psi_\R}
\eea
where ${{\mu'}^i_u} = \tilde\lambda'_i \Lambda_\hc$ with 
$\tilde\lambda'_i$ given in eq.\ (\ref{CKMmix}), and 
${{\mu}^i_d} = \tilde\lambda_i \Lambda_\hc$.  After rotating the
heavy and light fields to eliminate mass mixing, these lead to
magnetic moments involving just the light states.

%
\begin{figure}[t]
\hspace{-0.4cm}
\centerline{
\includegraphics[width=0.95\hsize]{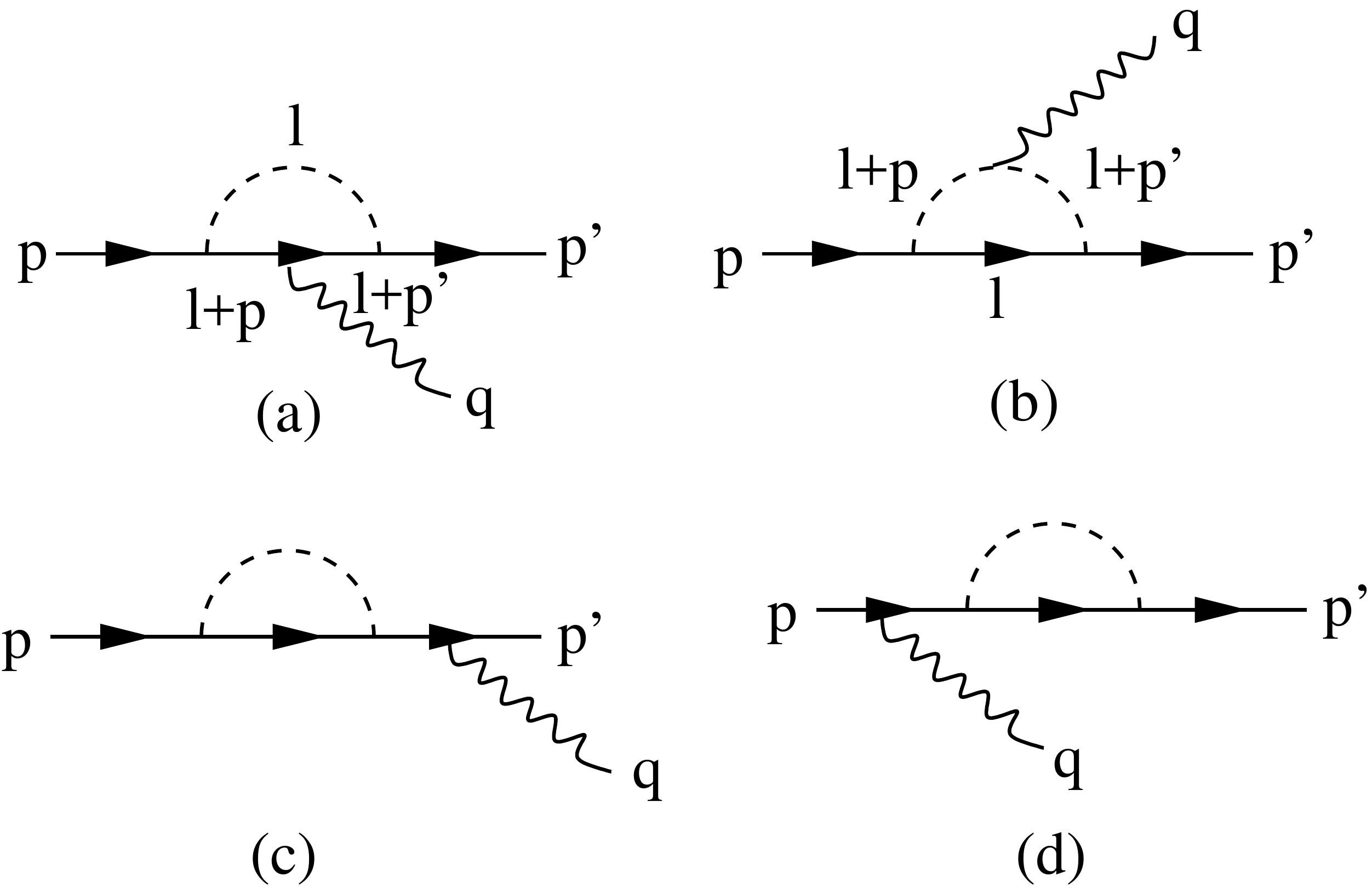}}
\caption{Contributions to elementary dark matter magnetic dipole
moment.}
\label{ssg}
\end{figure}
%

\section{DM dipole moment}
\label{DMDM}

We give details for the one-loop contribution to the magnetic 
dipole moment of the elementary dark matter particle $S$, 
with diagrams shown in fig.\ \ref{ssg}.  They
 turn out to be infrared convergent in the 
limit $m_\mu\to 0$ for the internal muon, hence we make this
simplification.  The 1PI diagrams (a,b) respectively give
\be
e|\lambda_2|^2\, 2\!\int_0^1 
\!\!\!dx\!\!\int_0^{1-x}\!\!\!\!\!\!\!\!\!\!dy\int\!\!\! {d^{\,4}l\over
(2\pi)^4}\,\bar u_{p'}{(\slashed{l}+\slashed{p'})\gamma^\mu(\slashed{l}+\slashed{p})
P_\R\over [(l + xp' + yp)^2 - M_a^2]^3} \,u_p\nn\\
\ee
\be
e|\lambda_2|^2\, 2\!\int_0^1 
\!\!\!dx\!\!\int_0^{1-x}\!\!\!\!\!\!\!\!\!\!dy\int\!\!\! {d^{\,4}l\over
(2\pi)^4}\,\bar u_{p'}{(2l-p-p')^\mu \slashed{l}
P_\R\over [(l - xp - yp')^2 - M_b^2]^3} \,u_p\nn\\
\ee
where, defining $u=x+y$,
\bea
	M_a^2 &=& (1-u)\,[ m_\phi^2 - u\, m_S^2]\nn\\
	M_b^2 &=& u\, [m_\phi^2 -(1-u)\,m_S^2]
\eea
As usual, one shifts the loop momentum variable to simplify the 
denominators.  Only the finite parts contribute to the 
dipole operator.
To extract the dipole moment, one makes the replacements
\bea
	P_\L\slashed{p}\gamma^\mu &\to& P_\L(\slashed{p}+\slashed{q})\gamma^\mu
	\to (P_\R m_S + P_\L \slashed{q})\gamma^\mu\nn\\
	\gamma^\mu\slashed{p'}P_\R &\to&
\gamma^\mu(\slashed{p}-\slashed{q})P_\R \to \gamma^\mu(m_S P_\L - 
\slashed{q} P_\R)
\eea
in diagram (a), using the Dirac equation (and similarly 
letting $\slashed{p'}\gamma^\mu\to m_S\gamma^\mu$,
$\gamma^\mu\slashed{p} \to m_S\gamma^\mu$), and keeping the terms
linear in $\slashed{q}$.  Both of these can be put into the form
$\gamma^\mu\slashed{q}$ up to terms going as $q^\mu$, that give a
vanishing contribution when contracted with the external photon
polarization vector.   These are directly related to the magnetic
moment since $-i\sigma_{\mu\nu}q^\nu \to \gamma^\mu\slashed{q}$ in 
this way.  

In diagram (b), we again use the Dirac equation to rewrite
$\slashed{p}P_\R \to m_S P_\L$, $P_\L\slashed{p'}\to m_S P_\R$, 
and reexpress $p^\mu,p'^\mu$ as
linear 
combinations of $(p+p')^\mu$ and $q^\mu$, of which only the former
contribute to the amplitude.  Using the Gordon identity, $(p+p')^\mu
\to i\sigma_{\mu\nu}q^\nu \to -\gamma^\mu\slashed{q}$. We find that
both diagrams (a,b) contribute with the same sign.  A check on this relative
sign is provided by gauge invariance: the coupling $\bar S\gamma^\mu S
A_\mu$ is forbidden, and must not be generated by loops.  The
divergent parts of diagrams (a,b) contribute with the same magnitude
and sign to the
vector current, and these cancel the contributions from (c,d).  Such a
cancellation would not occur if (a) and (b) came with opposite signs.

Defining $R = m_S^2/m_\phi^2$, 
the final expression for the magnetic moment (which is the coefficient
of $\bar u_{p'}\gamma^\mu \slashed{q} u_p$) is
\be
\mu_S = {e|\lambda_2|^2 \over 32\pi^2\,m_\phi^2 }\left(f_a(R) + f_b(R)\right)
\ee
with
\bea	
	f_a(R) &=& \int_0^1 du {u^2\over 1- Ru}\nn\\
	f_b(R) &=& \int_0^1 du{ u(1-u)\over 1 - R + Ru}
\eea
In the limit of light dark matter, $f_a(0) + f_b(0)
= 1/2$. 

\vfill
\bibliographystyle{apsrev}

\end{document}